\pdfoutput=1
\documentclass{rspublic}

\usepackage{amsmath,amssymb,amsfonts,amsthm,graphicx,dcolumn}
\usepackage{natbib}

\usepackage{bm}
\usepackage[mathscr]{eucal}
\usepackage{psfrag,theorem}
\usepackage{subfigure}
\usepackage{wrapfig}
\usepackage{color}
\include{graphix}
\usepackage{framed}
\usepackage{theorem}
\usepackage{enumitem}

\graphicspath{{figs_pdf/}}

\newcommand{\ai}{\alpha_{\mathrm{i}}}
\newcommand{\ar}{\alpha_{\mathrm{r}}}

\newcommand{\omi}{\omega_{\mathrm{i}}}
\newcommand{\omizero}{\omega_{\mathrm{i}0}}
\newcommand{\omitemp}{\omega_{\mathrm{i}}^{\mathrm{temp}}}
\newcommand{\omrtemp}{\omega_{\mathrm{r}}^{\mathrm{temp}}}
\newcommand{\omr}{\omega_{\mathrm{r}}}
\newcommand{\cg}{c_{\mathrm{g}}}

\newcommand{\am}{\alpha_{\mathrm{m}}}
\newcommand{\omim}{\omega_{\mathrm{i\,m}}^{\mathrm{temp}}}

\newcommand{\imag}{\mathrm{i}}
\newcommand{\mathe}{\mathrm{e}}

\newcommand{\partialim}{\partial_{\mathrm{i}}}
\newcommand{\partialr}{\partial_{\mathrm{r}}}

\newcommand{\smallparam}{\varepsilon}

\newcommand{\disp}{\mathcal{D}}

\newcommand{\xxi}{X_{\mathrm{i}}}
\newcommand{\xr}{X_{\mathrm{r}}}

\begin{document}
\title[Spatio-temporal instability]{An analytical connection between temporal and spatio-temporal growth rates in linear stability analysis}
\author[L. \'O N\'araigh and P. D. M. Spelt]{Lennon \'O N\'araigh$^1$ and Peter D. M. Spelt$^2$}
\affiliation{$^1$School of Mathematical Sciences, University College Dublin, Belfield, Dublin 4 
$^2$Laboratoire de M\'{e}canique des Fluides \& 
d'Acoustique, CNRS, Ecole Centrale Lyon, Ecully, France, and D\'{e}partement de M\'{e}canique, Universit\'{e} de Lyon 1, France}
\date{\today}

\label{firstpage}
\maketitle

\begin{abstract}{Absolute instability, Spatio-temporal linear stability, Orr--Sommerfeld equation}

We derive an exact formula for the complex frequency in  spatio-temporal stability analysis that is valid for arbitrary complex wave numbers.  The usefulness of the formula lies in the fact that it depends only on purely temporal quantities, which are easily calculated.  We apply the formula to two model dispersion relations: the linearized complex Ginzburg--Landau equation, and a model of wake instability.  In the first case, a quadratic truncation of the exact formula applies; in the second, the same quadratic truncation yields an estimate of the parameter values at which the transition to absolute instability occurs; the error in the estimate decreases upon increasing the order of the truncation.  We outline ways in which the formula can be used to characterize stability results obtained from purely numerical calculations, and point to a further application in global stability analyses.
\end{abstract}

\section{Introduction}
\label{sec:intro}
We consider linear stability analyses of general fluid flows, and begin by recalling the main steps of these, before introducing the motivation of the present work, thereby also introducing the necessary notation. We  first restrict ourselves to the linear stability analysis of a parallel flow $U_0(z)$ to two-dimensional  disturbances (extensions of the work to non-parallel flows and to three-dimensional disturbances are set out in \S\ref{sec:disc}). Such analysis is accomplished by solving the Orr--Sommerfeld (OS) equation~\citep{orr1907}. The solution of the equation at fixed parameter values (such as the Reynolds number $Re$) yields an eigenvalue problem that connects the wave number, the frequency, and the growth rate of the disturbance~\citep{DrazinReidBook}.  For two-dimensional disturbances, the perturbation streamfunction has the form $\psi(x,z,t)=\phi(z)\mathe^{\imag \left(\alpha x-\omega t\right)}$,
where $\alpha=\ar+\imag\ai$ and $\omega=\omr+\imag\omi$ are  complex numbers, $\ar$ is the (real) wave number, $\omr$ is the (real) frequency, and $\ai$ and $\omi$ are the spatial and temporal growth rates respectively; in this context, the OS equation reads
\begin{equation}
\imag\alpha\left(U_0(z)-c\right)\left(\partial_z^2-\alpha^2\right)\phi-\imag\alpha U_0''(z)\phi=Re^{-1}\left(\partial_z^2-\alpha^2\right)^2\phi,\qquad c=\omega/\alpha.
\label{eq:os0}
\end{equation}
 That the wavenumber $\alpha$ is complex indicates that the instability is spatio-temporal in nature~\citep{huerre90a}; in a \textit{temporal analysis}~\citep{DrazinReidBook}, $\ai=0$ and the eigenvalue problem is solved for $\omi$ and $\omr$.  This results in the following temporal {dispersion relations} for the temporal frequency and growth rate respectively: $\omrtemp=\omr(\ar,\ai=0)$, and $\omitemp=\omi(\ar,\ai=0)$.
The basic flow is called \textit{unstable} if the growth rate is positive for some $\ar$.  It is of interest further to classify unstable flows as  convectively or absolutely unstable: the flow is  convectively unstable if  initially localized pulses are amplified in at least one moving frame of reference but are damped in the laboratory frame; on the other hand, the flow is absolutely unstable if such pulses lead to growing disturbances in the entire domain in the laboratory frame.

\citet{Briggs1964} developed a method to classify unstable plasmas according to this dichotomy.  This criterion was extended to fluid dynamics by a number of authors~\citep{Gaster1968,Kupfer1987,Brevdo1988,huerre90a,Lingwood1997} 
The approach is based on a {\it local} eigenvalue analysis of the OS equation, which forms the key ingredient in global analyses: (i) a necessary condition for absolute instability is if the imaginary part of the frequency is positive at a saddle point in the complex $\alpha$-plane: $\omizero:=\omi(\alpha_0)>0$, where $\alpha_0$ solves $d\omega/d\alpha=0$; (ii)  to obtain a sufficient condition for absolute instability, the saddle point $\alpha_0$ in the complex $\alpha$-plane must be the result of the coalescence of spatial branches that originate from opposite half-planes at a larger and positive value of $\omi$ (this coalescence or `pinching' of the spatial branches is accompanied by the formation of a cusp at $\omizero$ in the complex $\omega$ plane).
Typically, the saddle-point and Briggs criteria are checked using a fast numerical eigenvalue solver~\citep{Boyd,TrefethenBook}.  Our purpose here is not to supplant such effective computations,
but rather to develop an analytical connection between spatio-temporal and
temporal growth rates to assist in understanding these numerical results. A second
motivation is the possible extension of this analytical approach to globally-unstable
flows. We discuss these motivations in more detail now.

Our purpose within a local description is to develop an analytical connection between spatio-temporal and temporal growth rates. Such a connection is desirable for a number of reasons.
First, in flows where several temporal modes are unstable, 
the interpretation of the results of a spatio-temporal stability analysis (obtained numerically, for instance)  becomes difficult, as does the mere determination of the convective/absolute (C/A) transitions~\citep{Suslov06}. In this case, $\omega$-plots of the distinct spatio-temporal modes consist of complicated interpenetrating surfaces, whose domain is the entire complex $\alpha$-plane. On the other hand, many theoretical tools exist for characterizing temporal instabilities, including energy-budget analyses (as reviewed by \citet{Boomkamp1996}) and asymptotic analytical solutions~\citep{DrazinReidBook}, so the governing physical mechanisms for growth of temporal modes is often well documented and competing temporal modes can easily be distinguished. A direct link with temporal modes could therefore enable one to distinguish the competing spatio-temporal modes. Furthermore,  in parametric studies involving large parameter spaces (e.g., thermal boundary layers or two-phase flows), wherein the C/A transition curves are plotted as a function of the flow parameters, the parametric dependence of the C/A transition curves may be difficult to analyse.  For example, for a system with a two-dimensional parameter space $(\mu_1,\mu_2)$, the curve demarcating the transition between convective and absolute instability is given by the generic formula $\mu_2=f(\mu_1)$.  Here, knowledge of the physical properties of the system (in particular the familiar temporal stability properties) may be beneficial (if not crucial) to deduce the function $f(\cdot)$.

Although the main focus in this study is on an analytic connection between spatio-temporal and temporal local stability growth rates, the results are also of interest in {\it global} stability analyses. There, the base state can be determined numerically or otherwise (using either an unperturbed solution or a time-averaged perturbed result), in terms of the streamwise coordinate. Assuming this to evolve slowly with the streamwise coordinate the saddle-node frequency $\omizero$ can then be determined as a function of the real scaled streamwise coordinate, $X$. But  in order to determine global modes (detailed criteria are recalled briefly in \S~\ref{sec:disc}), $\omizero (X)$ is required off the real axis, for which analytic continuation is used. In other words, rather than a connection between growth rates for complex and real wave numbers, one between growth rates for complex and real spatial coordinates is needed. An example of 
wherein analytical continuation is used for this purpose is the study of~\citet{Hammond1997}. The formulation proposed herein in the context of a local analysis is readily reformulated for use in this setting for a global analysis, as outlined at the end of this paper, and offers the possibility of developing a combination of both.

This work is organized as follows. We derive an exact formula connecting the spatio-temporal growth rate with purely temporal quantities in \S~\ref{sec:exact}.  We discuss the two lowest-order truncations of this formula in \S\ref{sec:limiting}: Gaster's formula, and a situation in which the temporal quantities depend only quadratically on the wavenumber $\ar$.  This enables the formulation of condition (i) into a succinct equation encoding the competition between \textit{in-situ} growth and convective effects.  In \S\ref{sec:singular} we describe the singularities that typically occur in the dispersion relation, and discuss how these can hamper the convergence of our formula.  In \S\ref{sec:numerics} we apply our results to a simple OS analysis involving wake instability.  In \S\ref{sec:disc} we summarize our arguments and describe further applications of the exact formula and its quadratic approximation, including in a global stability analysis.

\section{An exact formula for the complex frequency, derived from purely temporal quantities}

\label{sec:exact}

The development of the formula starts with the assumption that an eigenvalue analysis of the OS equation yields a complex frequency $\omega$ that depends on the streamwise wave number $\alpha$ as an analytic (holomorphic) function, $\omega=\omr(\ar,\ai)+\imag \omi(\ar,\ai)$,
%
%
%
where we have fixed the other system parameters (e.g. Reynolds number).  The domain $D$ of analyticity is an open simply-connected subset of $D\subset\mathbb{C}$, containing a part of the real line.  Details concerning  the domain are discussed further below.
As a consequence of the analyticity of $\omega(\alpha)$, we have the following Cauchy--Riemann conditions:
\begin{subequations}
\vspace{-0.25in}
\begin{center}
\begin{tabular}{p{0.4\textwidth}p{0.4\textwidth}}
{\begin{align}
&\frac{\partial\omr}{\partial \ar}=\frac{\partial\omi}{\partial\ai},
\label{eq:cr1}
\end{align}}
&
{\begin{align}
& \frac{\partial\omr}{\partial \ai}=-\frac{\partial\omi}{\partial\ar}.
\label{eq:cr2}
\end{align} }
\end{tabular}
\end{center}
\vspace{-0.2in}
\end{subequations}
We use Equations~\eqref{eq:cr1} and~\eqref{eq:cr2} to derive an expression for $\omi(\ar,\ai)$ as a function of the purely temporal stability properties.

We work at a fixed value of $\ar$ in Equation~\eqref{eq:cr1} and identify the group velocity $\cg(\ar)=(\partial\omr/\partial\ar)_{(\ar,0)}$.
%
%
Furthermore, we Taylor-expand $\partial\omr/\partial\ar$ into the region $D$:
\begin{equation}
\frac{\partial\omr}{\partial\ar}=\sum_{n=0}^\infty \frac{1}{n!}c_{\mathrm{g}n}(\ar)\ai^n,\qquad
c_{\mathrm{g}n}(\ar)=\frac{\partial^n}{\partial\ai^n}\frac{\partial\omr}{\partial\ar}\bigg|_{\ai=0}.
\end{equation}
This amounts to a complex Taylor expansion centred at $(\ar,0)$, and is therefore valid on a disc of radius $R$ contained entirely in $D$.  The magnitude of the radius of convergence $R$ is discussed below.
Each term in this Taylor expansion is available from a purely temporal analysis.  We describe this process for the first few terms (for brevity, we use $\partialim:=\partial/\partial\ai$ and $\partialr:=\partial/\partial\ar$).
At $n=0$ we have $c_{\mathrm{g}0}=\partialr\omr|_{\ai=0}=\cg(\ar)$, the group velocity.  At $n=1$ we obtain
\begin{equation}
c_{\mathrm{g}1}=\partialim\partialr\omr=\partialr\partialim\omr\overset{\mathrm{C.R.}}=-\partialr\partialr\omi\overset{\ai=0}=-\frac{d^2\omitemp}{d\ar^2},
\end{equation}
where $\omitemp$ is the purely temporal growth rate.
At $n=2$ we have
\begin{equation}
c_{\mathrm{g}2}=\partialim\partialim\partialr\omr=\partialim\partialr\partialim\omr\overset{\mathrm{C.R.}}=-\partialim\partialr\partialr\omi=-\partialr\partialr\partialim\omi\overset{\mathrm{C.R.}}=-\partialr\partialr\partialr\omr
\overset{\ai=0}=-\frac{d^2\cg}{d\ar^2}.
\end{equation}
At $n=3$ we compute
\begin{eqnarray}
c_{\mathrm{g}3}&=&\partialim\partialim\partialim\partialr\omr=\partialim\partialim\partialr\partialim\omr\overset{\mathrm{C.R.}}=-\partialim\partialim\partialr\partialr\omi=-\partialim\partialr\partialr\partialim\omi\nonumber\\
&\overset{\mathrm{C.R.}}=&-\partialim\partialr\partialr\partialr\omr
=-\partialr\partialr\partialr\partialim\omr
\overset{\mathrm{C.R.}}=\partialr\partialr\partialr\partialr\omi
\overset{\ai=0}=\frac{d^4\omitemp}{d\ar^4}.
\end{eqnarray}
Similarly, at $n=4$ we have
\begin{eqnarray}
c_{\mathrm{g}4}&=&\partialim\partialim\partialim\partialim\partialr\omr=\partialim\partialim\partialim\partialr\partialim\omr\overset{\mathrm{C.R.}}=-\partialim\partialim\partialim\partialr\partialr\omi=
-\partialim\partialim\partialr\partialr\partialim\partial\omi\nonumber\\
&\overset{\mathrm{C.R.}}=&-\partialim\partialim\partialr\partialr\partialr\omr=
-\partialim\partialr\partialr\partialr\partialim\omr\overset{\mathrm{C.R.}}=\partialim\partialr\partialr\partialr\partialr\omi=
\partialr\partialr\partialr\partialr\partialim\omi\nonumber\\
&\overset{\mathrm{C.R.}}=&\partialr\partialr\partialr\partialr\partialr\omr
\overset{\ai=0}=\frac{d^4\cg}{d\ar^4}.
\end{eqnarray}
We deduce the higher-order terms by inspection.  Assembling these results, we obtain  an expression for $\partial\omr/\partial\ar$, extended into the complex plane:
\begin{equation}
\frac{\partial\omr}{\partial\ar}\bigg|_{(\ar,\ai)}=\sum_{n=0}^\infty \frac{(-1)^n}{(2n)!}\frac{d^{2n}\cg}{d\ar^{2n}}\ai^{2n}+\sum_{n=0}^\infty \frac{(-1)^{n+1}}{(2n+1)!}\frac{d^{2n+2}\omitemp}{d\ar^{2n+2}}\ai^{2n+1}.
\label{eq:omr_taylor}
\end{equation}
Next, we use the Cauchy--Riemann condition~\eqref{eq:cr1} and connect the Taylor expansion in Equation~\eqref{eq:omr_taylor} to an expression for $\omi(\ar,\ai)$, valid in the entire complex half-plane $\ar>0$.  We have
\begin{equation}
\frac{\partial\omi}{\partial\ai}\overset{\mathrm{C.R.}}=\frac{\partial\omr}{\partial\ar}
=\sum_{n=0}^\infty \frac{(-1)^n}{(2n)!}\frac{d^{2n}\cg}{d\ar^{2n}}\ai^{2n}+\sum_{n=0}^\infty \frac{(-1)^{n+1}}{(2n+1)!}\frac{d^{2n+2}\omitemp}{d\ar^{2n+2}}\ai^{2n+1}.
\end{equation}
We integrate this relation along a line perpendicular to the real axis, from $(\ar,0)\rightarrow (\ar,\ai)$ ($\ai$ is arbitrary, i.e. $|\ai|$ is not necessarily small).  The result is
\begin{equation}
\omi(\ar,\ai)=\omitemp(\ar)+\sum_{n=0}^\infty \frac{(-1)^n}{(2n+1)!}\frac{d^{2n}\cg}{d\ar^{2n}}\ai^{2n+1}
+\sum_{n=0}^\infty \frac{(-1)^{n+1}}{(2n+2)!}\frac{d^{2n+2}\omitemp}{d\ar^{2n+2}}\ai^{2n+2}.
\label{eq:omi_taylor}
\end{equation}
This is an exact formula for the imaginary part of the complex frequency, which depends only on purely temporal quantities.  Mathematically, this is a trivial statement: since the complex frequency is analytic, 
analytic continuation implies that its behaviour on the real line completely determines its behaviour on $D$.  Nevertheless, this formula will help us to create a practical means of predicting  the threshold for absolute instability.

The series~\eqref{eq:omi_taylor} amounts to a complex Taylor series centred at the point $(\ar,0)$ and therefore converges inside a disc of radius $R$.  The radius $R$ is the minimum distance from the point $(\ar,0)$ to the nearest singularity of $\omega(\alpha)$.  Care must be taken that the point of interest (e.g. the saddle point of $\omega(\alpha)$) lies inside this disc.   A detailed discussion of the radius of convergence of Equation~\eqref{eq:omi_taylor}  is given in \S\ref{sec:singular}.

\section{Low-order truncations and an approximate criterion for the C/A transition}
\label{sec:limiting}

In this section we investigate the implications of truncating Equation~\eqref{eq:omi_taylor} at linear order, and at quadratic order.  At linear order, the spatial analysis of Gaster is recovered; at quadratic order, the complex Ginzburg--Landau model is obtained, albeit with further understanding, in the form of a `balance condition' for the onset of absolute instability.

\subsection{Relation to the analysis of Gaster}
\label{sec:limiting:gaster}

The Gaster transformation~\citep{Gaster1962} concerns purely \textit{spatial} instabilities, which correspond to setting $\omi=0$ in the OS eigenvalue analysis, and computing the resulting spatial growth rate $\ai$ as a function of $(\omr,\ar)$.  Practically, this corresponds to a localized disturbance in the flow that oscillates at a characteristic frequency $\omr$, and grows downstream of the disturbance at a (spatial) rate $-\ai$.  Spatial growth therefore corresponds to negative values of $\ai$.  Using the fact that the general complex frequency $\omega(\alpha)$ is an analytic function, Gaster integrated the resulting Cauchy--Riemann conditions along a straight-line path of small length, $(\ar,0)\rightarrow (\ar,\ai)$, with $|\ai|\ll 1$.  The resulting integrands are thus regarded as constant and, setting $\omi=0$, Gaster obtained the following formula for the spatial growth rate:
\begin{equation}
\ai=-\omitemp\Big\slash\frac{d\omrtemp}{d\ar}.
\label{eq:gaster0}
\end{equation}
This equation provides a precise connection between the spatial growth rate and more familiar temporal quantities.  Nevertheless, it is limited, in the sense that it applies only to spatial growth (as opposed to fully spatio-temporal growth) and is valid only for small values of $|\ai|$.

Equation~\eqref{eq:gaster0} can be recovered directly from the approach in \S\ref{sec:exact}. Consider Equation~\eqref{eq:omi_taylor} with $|\ai|\ll 1$, such that only the lowest-order term gives a contribution to the equation.  The result is
\begin{equation}
\omi(\ar,\ai)\sim \omitemp(\ar)+\cg(\ar)\ai,\qquad |\ai|\rightarrow 0.
\label{eq:gaster_reduce1}
\end{equation}
On a spatial branch, we have $\omi=0$, hence Equation~\eqref{eq:gaster_reduce1} becomes $$\ai\sim -\omitemp(\ar)/\cg(\ar),$$ as $|\ai|\rightarrow 0$, 
which is precisely the formula of~\citet{Gaster1962} (Equation~\eqref{eq:gaster0}) for spatial modes.

\subsection{The quadratic approximation }
\label{sec:limiting:quadratic}

We also derive conditions for the C/A transition based on a second-order truncation of the series dispersion relation~\eqref{eq:omi_taylor} -- the quadratic approximation.
%
%
%
%
%
%
%
%
%
%
%
In this truncation,  the Taylor series~\eqref{eq:omi_taylor} reduces to
\begin{equation}
\omi(\ar,\ai)=\omitemp(\ar)+\cg(\ar)\ai-\tfrac{1}{2}\frac{d^2\omitemp}{d\ar^2}\ai^2.
\label{eq:omi_approx}
\end{equation}
The necessary saddle-point condition for an absolute instability (as reviewed in \S\ref{sec:intro}) is $\partial\omi/\partial\ar=\partial\omi/\partial\ai=0$; using Equation~\eqref{eq:omi_approx}, these conditions amount to
\begin{equation}
\frac{d\omitemp}{d\ar}+\frac{d\cg}{d\ar}\ai-\tfrac{1}{2}\frac{d^3\omitemp}{d\ar^3}\ai^2=0,
\qquad
\cg(\ar)-\frac{d^2\omitemp}{d\ar^2}\ai=0.
\label{eq:ai_linear}
\end{equation}
%
%
Taking the second condition, we get
\begin{equation}
\ai=\cg(\ar)\bigg\slash \frac{d^2\omitemp}{d\ar^2}.
\end{equation}
Substitution into the first condition yields
\begin{equation}
\cg(\ar)\frac{d\cg}{d\ar}=-\frac{d\omitemp}{d\ar}\frac{d^2\omitemp}{d\ar^2}+\tfrac{1}{2}\cg^2(\ar)\left(\frac{d^3\omitemp}{d\ar^3}\bigg\slash\frac{d^2\omitemp}{d\ar^2}\right).
\label{eq:saddle}
\end{equation}
(Note that third-order derivatives appear in this calculation -- but only through the process of finding the saddle-point location, and not in the series expansion of the dispersion relation.)

The existence of a real root of Equation~\eqref{eq:saddle} is a necessary condition for a saddle point to occur.  
Once the root of Equation~\eqref{eq:saddle} is extracted, we derive a condition for the $\omi$ to vanish at the saddle point, as this is the sign of the transition to absolute instability.  Referring to Equation~\eqref{eq:omi_approx}, $\omi$ vanishes at the saddle when
\begin{equation}
-\frac{d^2\omitemp}{d\ar^2}\bigg|_{\ar^*}\omitemp(\ar^*)=\tfrac{1}{2}\cg^2(\ar^*).
\label{eq:saddle_sign}
\end{equation}
where $\ar^*$ is the root of Equation~\eqref{eq:saddle}.  Knowledge of the saddle-point location $\ar^*$, together with the dispersion relation~\eqref{eq:omi_approx} yield further information that can be used to verify if the saddle point arises as a result of the coalescence of spatial branches ramifying into different halves of the complex $\alpha$-plane (i.e. the necessary and sufficient conditions for the onset of absolute instability).
Finally, we note that the dispersion relation for the model linear  complex Ginzburg--Landau equation is quadratic in the wavenumber~\citep{huerre2000,Suslov2004}, and no error is incurred in this case in applying the quadratic truncation of the series dispersion relation~\eqref{eq:omi_taylor}.
%
%
%
%
%
%
%
%
%
%

\subsection{Higher-order approximations}

For certain flows, a higher-order truncation of the series~\eqref{eq:omi_taylor} may be required, in order to capture the multiplicity of saddle points that can contribute to the spatio-temporal growth in certain anomalous situations~\citep{Brevdo1999}.  A straightforward extension of the analysis in \S\ref{sec:limiting}\ref{sec:limiting:quadratic} applies.  The reader is referred to the Supplementary Material for details of an application of the series expansion of the dispersion relation for free-surface flow on an inclined plane~\citep{Brevdo1999}.


An advantage of the framework developed in this section (compared with the direct approach) concerns the insight given by Equation~\eqref{eq:saddle_sign} into the balance of competing effects that brings about absolute instability.
Intuitively, we may think of the condition~\eqref{eq:saddle_sign} as a competition between \textit{in-situ} growth on the left and convection of the disturbance on the right.  Absolute instability can set in only if the \textit{in-situ} growth equals or exceeds the tendency of the parallel flow to convect disturbances downstream.  Similar advantages carry over to situations where high-order truncations are required: the existence of multiple $\ai$-roots in the polynomial truncation of the criterion $d\omi/d\ai=0$ corresponds exactly to multiple branches in the spatio-temporal growth rate in a moving frame~\citep{Brevdo1999}.
Finally, we note that although a `quadratic approximation' was pursued in earlier works, the context was different: either in relation to a numerical, iterative procedure for determining the location of the saddle point (which anyway fails in the presence of multiple saddle points)\citep{Deissler1987,Monkey1988}, or in the context of asymptotic theories concerned with reduction of a complicated model to a Complex Ginzburg--Landau equation~\citep{Suslov2004}.
None of these works contains the explicit condition linking the C/A transition to the temporal analysis which is a key result of this section.

\section{Convergence of series for $\omi$ in the presence of a bestiary of singularities} 
\label{sec:singular}

In this section we review the different kinds of singularity that can occur in the dispersion relation $\omega=\omega(\alpha)$, and discuss the implications of such possible non-analytic behaviour for the convergence of our formula~\eqref{eq:omi_taylor}.  Here, the `impulse response function' refers to the solution of the spatiotemporal problem
\begin{multline}
\left[U_0(z)\partial_x+\partial_t\right]\left(\partial_z^2+\partial_x^2\right)\psi(x,z,t)-U_0''(z)\partial_x\psi(x,z,t)=Re^{-1}\left(\partial_z^2+\partial_x^2\right)^2\psi(x,z,t),\\
\psi(x,z,t=0)=\delta(z)\delta(x).
\label{eq:impulse}
\end{multline}
%
%
%
Equation~\eqref{eq:impulse} is typically solved using a Laplace-Fourier decomposition; the dispersion relation of the resulting eigenvalue problem (the OS equation) determines the shape of the impulse response.



\textit{1.  Branch cuts in unconfined flows:}  In so-called \textit{unconfined flows}, the dispersion relation $\omega=\omega(\alpha)$ may possesses branch cuts along the imaginary axis $\ar=0$~\citep{huerre85,Lingwood1997}.  In such systems (e.g. mixing layers, flow past obstacles), both the streamwise and normal dimensions of the fluid container extend to infinity (labelled by coordinates $x$ and $z$ respectively), and the disturbance streamfunction decays to zero as $|z|\rightarrow\infty$.  Then, the asymptotic ($|z|\rightarrow\infty$) OS equation reads
\[
\imag\alpha\left(U_\infty-c\right)\left(\partial_z^2-\alpha^2\right)\phi=Re^{-1}\left(\partial_z^2-\alpha^2\right)\phi,\qquad c=\omega/\alpha,
\]
for a symmetric base state $U_0(z)$ with 
\[
U_\infty=\lim_{z\rightarrow\pm\infty}U_0(z), \text{ and } \lim_{z\rightarrow\pm\infty}U_0''(z)=0.
\]
This asymptotic solution possesses an inviscid mode $\phi\sim\mathe^{-\mathrm{sign}(\ar)\alpha z}$ and a viscous mode $\phi\sim\mathe^{-\mathrm{sign}(\gamma_\mathrm{r})\gamma z}$, where $\gamma=\sqrt{\alpha^2+\imag\alpha Re(U_\infty-c)}$.  The inviscid mode induces a branch cut along the imaginary axis~\citep{huerre85}, while the viscous mode induces hyperbolic branch cuts in the $\alpha$-plane~\citep{Aships1990,Schmid2001}.    All of these branch cuts (or `continuous spectra'~\citep{Schmid2001}) contribute to the contour integral associated with the impulse-response function.  The contour integral of the impulse-response function therefore possesses two contributions: one from the zeros of the dispersion relation $\omega(\alpha)=0$ (`poles'), and the other from the continuous  spectrum.   However, to diagnose absolute instability, it suffices to consider the discrete part, as the continuous spectrum can produce temporal growth only when the discrete part is absolutely unstable~\citep{Lingwood1997}.

\textit{2.  Discrete poles along the imaginary axis in the complex $\alpha$-plane:}  \citet{Healey2007,Healey2009} has shown that confining an \textit{inviscid} version of the flow described in (2) between two plates at $z=\pm H$ causes the character of the singularity along the axis $\ar=0$ to switch from a continuous branch cut to a set of discrete poles.  For, the inviscid OS (Rayleigh) equation
\[
\imag\alpha\left(U_0(z)-c\right)\left(\partial_z^2-\alpha^2\right)\phi-\imag\alpha U_0''(z)\phi=0,
\]
can be re-written as
\[
 \left(\partial_z^2-\alpha^2\right)\phi=\frac{\imag\alpha U_0''(z)\phi}{\imag\alpha U_0(z)-\imag \omega},
\]
such that the singularity $\omega=\infty$ corresponds to $(\partial_z^2-\alpha^2)\phi=0$.  This limiting equation  satisfies the appropriate boundary conditions for a confined flow (no-penetration at the walls where confinement is enforced)  whenever $\phi$ can be made into an oscillatory function, that is, when $\alpha=\pm n\pi\imag/H$, with $n=1,2,\cdots$.  These poles induce a series of saddle points near the imaginary axis.  These saddle points can dominate over the saddle point associated with the analogous unconfined flow for sufficiently small $H$.  Such saddles can even pinch (according to the Briggs criterion) and thus produce absolute instability~\citep{Healey2007,Healey2009}, leading to the conclusion that a confined flow is `more absolutely unstable' than its unconfined analogue.  

Since the introduction of viscosity to the problem does not change the large-$\omega$ form of the eigenvalue problem, it is expected that such poles will persist in the viscous case (with some modification of the large-$\omega$ streamfunction arising from the different boundary conditions in the viscous case)~\citep{Healey2012}.  This contention is supported by numerical evidence (\S\ref{sec:numerics}).

\textit{3.  Further branch cuts in the complex $\alpha$-plane:}
The generic dispersion relation $\disp(\alpha,\omega)=0$ of the OS equation naturally evokes consideration of the multivariable function $\disp(\alpha,\omega)$ itself.  This is an analytic function in each of its variables~\citep{Gaster1968}.  Moreover, $\partial \disp/\partial\omega$ has at least one zero~\citep{Gaster1968}.  In the neighbourhood of this point (labelled $(\alpha_0,\omega_0)$), $\disp(\alpha,\omega)$ can be expanded as
\begin{multline*}
\disp(\alpha,\omega)=\disp(\alpha_0,\omega_0)+\disp_\alpha(\alpha_0,\omega_0)(\alpha-\alpha_0)+\tfrac{1}{2}\disp_{\omega\omega}(\alpha_0,\omega_0)(\omega-\omega_0)^2\\
+\text{Higher-order terms},
\end{multline*}
where the subscripts $\alpha$ and $\omega$ indicate partial differentiation.  If the points $(\alpha,\omega)$ and $(\alpha_0,\omega_0)$ satisfy the dispersion relation $\disp=0$, then this equation can be recast as
\begin{equation}
\omega=\omega_0+\left(-\frac{\disp_\alpha(\alpha_0,\omega_0)}{\disp_{\omega\omega}(\alpha_0,\omega_0)}\right)^{1/2}\left(\alpha-\alpha_0\right)^{1/2},
\label{eq:riemann_surf}
\end{equation}
The dispersion relation $\omega(\alpha)$ is therefore non-differentiable at $\alpha_0$ and possesses a branch cut along a line emanating from the point $\alpha_0$.  Equivalently, one may regard the dispersion relation as being multi-valued, taking distinct values on the two distinct sheets of the Riemann surface described by Equation~\eqref{eq:riemann_surf}.

Such singularities can be problematic in a number of ways.  First, they limit the radius of convergence of the series dispersion relation~\eqref{eq:omi_taylor}, thus inhibiting the description of a dynamically-relevant saddle point via this route.  Examples of this kind are found in the works of~\citet{Brevdo1988,Juniper2006}.
Even graver, if a saddle point of the dispersion relation is sufficiently close the branch cut~\eqref{eq:riemann_surf}, the branch cut can prevent the saddle point from pinching, such that the Briggs criterion is not satisfied (see \S\ref{sec:intro}).  Equivalently, in such cases, it is not possible to deform the contour in the impulse-response integral into a steepest-descent path without enclosing the singular point $\alpha_0$.  Such saddles do not therefore generate absolute instability~\citep{Lingwood1997}.

%
It is clear from this list that knowledge of the global topography of the dispersion relation is required to determine absolute instability (on this point, see also~\cite{Lingwood1997}).  Equally, such knowledge is necessary to determine when Equation~\eqref{eq:omi_taylor} can be used with confidence.  Since Equation~\eqref{eq:omi_taylor} is the imaginary part of a complex-valued power series, it is valid in the disc of convergence of this complex-valued series.  The outermost radius  $R$ of this disc is given by
%
\begin{multline}
R=\text{Distance between the point of interest }(\ar,0)\\\text{and the nearest singularity in }\omega(\alpha);
\label{eq:roc0}
\end{multline}
a necessary and sufficient condition for the validity of Equation~\eqref{eq:omi_taylor} at the point $(\ar,\ai)$ is thus $|\ai|<R$.  
In practice, the requirement that the global topography of the dispersion relation $\omega(\alpha)$  be known will not limit the potential uses of the local formula~\eqref{eq:omi_taylor}:  since we envisage that our formula should be applied to the in-depth analysis of numerical results (for energy-budget analyses and the characterization of mode competition and C/A transition curves), this does not result in any loss of relevance for the formula~\eqref{eq:omi_taylor}.

\section{Application of the formula to a canonical fluid instability}
\label{sec:numerics}

The quadratic approximation holds true exactly only for model dispersion relations such as the linearized complex Ginzburg--Landau equation (\S~\ref{sec:limiting:quadratic}). In more practical systems, the exact series expansion (\ref{eq:omi_taylor}) would have to converge quickly for the quadratic approximation to be useful. The convergence properties have already been discussed in general terms in \S\ref{sec:singular}; here, we discuss convergence for the particular model (dimensionless) velocity field
\begin{equation}
U_0(z)=1-\Lambda+2\Lambda\big\{1+\sinh^{2N}\left[z\sinh^{-1}(1)\right]\big\}^{-1},\qquad \Lambda<0,
\label{eq:uzero}
\end{equation}
where $\Lambda$ and $N$ are dimensionless parameters, and $-\infty<z<\infty$.  Equation~\eqref{eq:uzero} models the steady wake profile generated by flow past a bluff body.  The quantity $\Lambda=(U_\mathrm{c}-U_\mathrm{max})/(U_\mathrm{c}+U_\mathrm{max})$ is the velocity ratio, where $U_\mathrm{c}$ is the wake centreline velocity and $U_\mathrm{max}$ is the maximum velocity.  Furthermore, $N$ is the shape parameter, which controls the ratio between the mixing-layer thickness and the width of the wake.  It ranges from $N=1$ (the `$\mathrm{sech}^2$ wake') to $N=\infty$, a `top-hat wake' bounded by two vortex sheets~\citep{Monkey1988}.
The base state~\eqref{eq:uzero} is absolutely unstable~\citep{Monkey1988} (this serves as a model test case for our purposes; for subsequent work on cylinder wakes, see~\citet{Pier2002}).  The associated OS equation possesses three independent parameters, $(\Lambda,N,Re)$, where $Re$ is the Reynolds number.

\subsection{Numerical method}
\label{sec:numerics:method}

Practical numerical methods should be confirmed with a notional \textit{ideal method}, which would preserve 
the branch-cut singularity in the dispersion relation $\omega(\alpha)$ along the imaginary axis, associated with the unbounded domain in Equation~\eqref{eq:uzero}.  
%
%
%
We  considered (1) a standard Chebyshev collocation method using Chebyshev polynomials as the basis functions, wherein confinement is introduced at $z=\pm H$, such that $\phi(\pm H)=\phi'(\pm H)=0$; we also considered (2) a Chebyshev collocation method equipped with basis functions appropriate to an infinite domain~\citep{Boyd}.
The streamfunction for Method (2) is expanded in terms of basis functions as follows:
\begin{equation}
\phi(z)=\sum_{n=0}^{N_C}a_n TB_n(z),\qquad TB_n(y)=\cos\left(n\,\mathrm{arccot}\left(\frac{y}{L}\right)\right),
\label{eq:chebydef}
\end{equation}
where $L$ is an adjustable map parameter (typically chosen to be unity, e.g.~\citep{Boyd})  and $y\in(-\infty,\infty)$.  Each rational basis function $TB_n(y)$ possesses the asymptotic behaviour $TB_n(y)\sim 1$ as $|y|\rightarrow\infty$.  The $a_n$'s are determined (along with the eigenvalue $-\imag\alpha c$) by substitution of Equation~\eqref{eq:chebydef} into the OS equation, evalution at collocation points, and solution of the resulting (finite) matrix eigenvalue problem.  Convergence of the numerical method is obtained by increasing $N_C$ until no change in the eigenvalue is observed (to within working precision).
One would expect this second approach to capture the continuous singularity along the imaginary axis.
%

Methods (1) and (2) yield identical results with respect to a purely temporal stability analysis: we have verified the correctness of these methods by computing the critical Reynolds number for the onset of (convective) instability: $Re_\mathrm{c}(N=1,\Lambda=-1)=1.8$, and $Re_\mathrm{c}(N=2,\Lambda=-1)=1.9$, in agreement with the existing literature~\citep{Monkey1988}. 
Not surprisingly, given the discussion in \S\ref{sec:singular}, we have found that method (1) produces a sequence of equally-separated poles in the dispersion relation along the imaginary-wavenumber axis.  
The value of $H$ is chosen as $H=H_\mathrm{c}$; here $H_\mathrm{c}$ is large in the sense that the linear-stability results are the same for $H=H_\mathrm{c}$ and $H>H_\mathrm{c}$.     The only exception is near the imaginary axis, where confinement features occur; these features are well separated from the physical saddle points in the dispersion relation and do not affect the stability results (non-physical singularities and saddle points are identified as they move around upon changing the amount of confinement).   In this work, we take $H_\mathrm{c}=8$.
On the other hand -- rather surprisingly -- we have found that method (2) does not produce a continuous spectrum in the dispersion relation along the imaginary-wavenumber axis, but rather a sequence of poles: even with the basis functions appropriate for the infinite domain, the numerical solution artificially  interprets the problem as though its geometry were confined.

\subsection{Results}
\label{sec:numerics:results}

\begin{figure}[htb]
\centering
\subfigure[]{\includegraphics[width=0.32\textwidth]{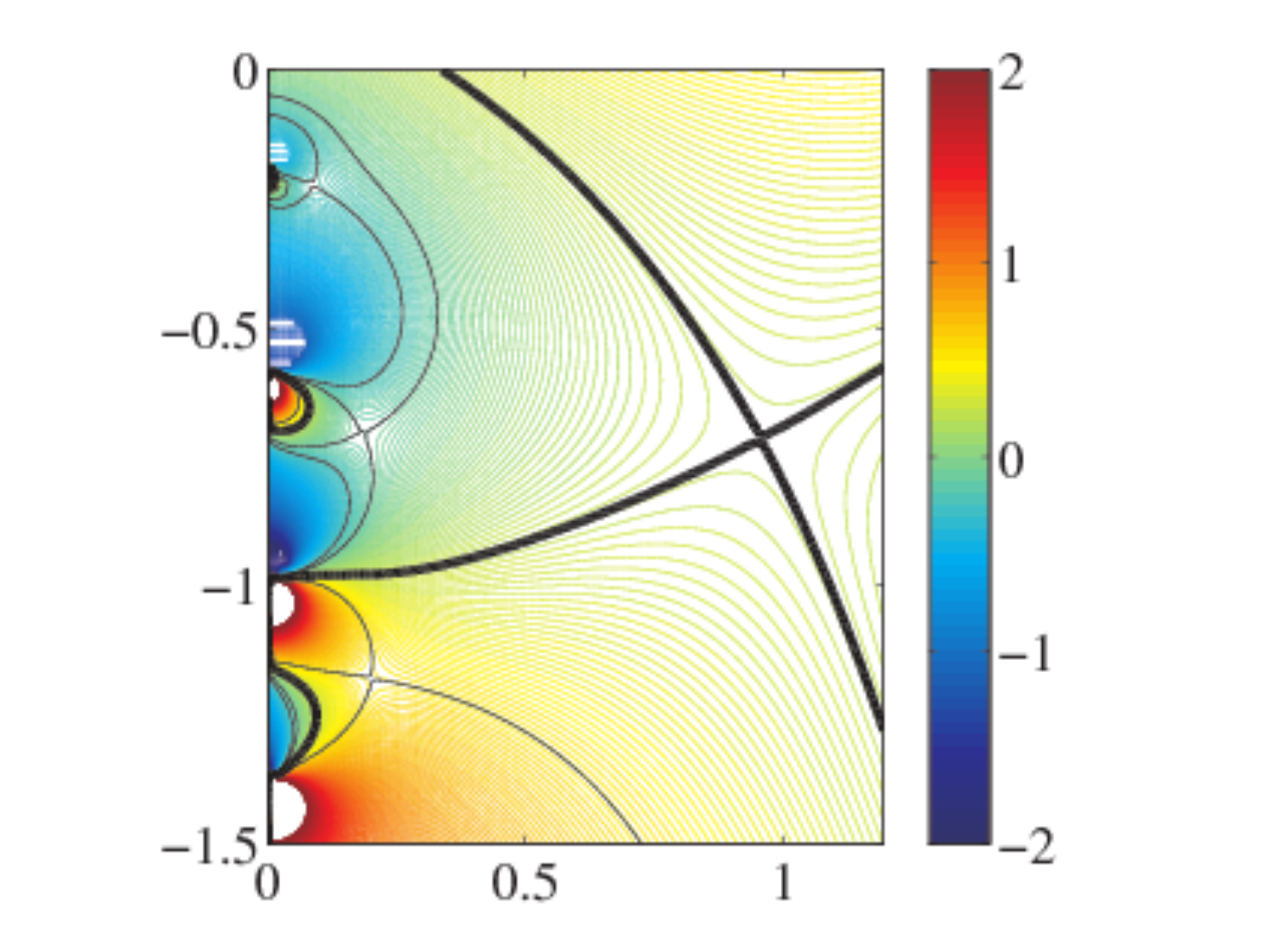}}
\subfigure[]{\includegraphics[width=0.32\textwidth]{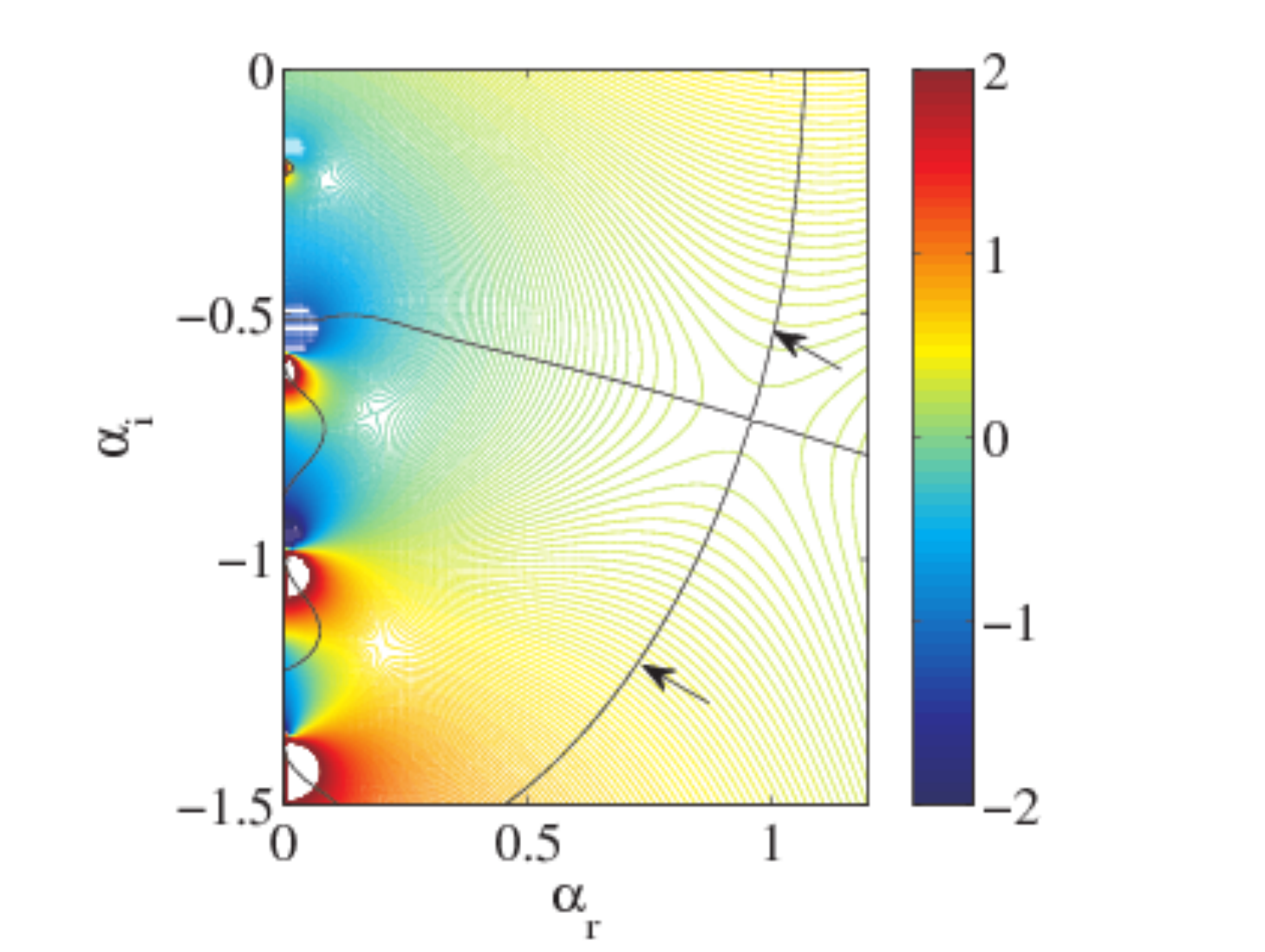}}
\subfigure[]{\includegraphics[width=0.32\textwidth]{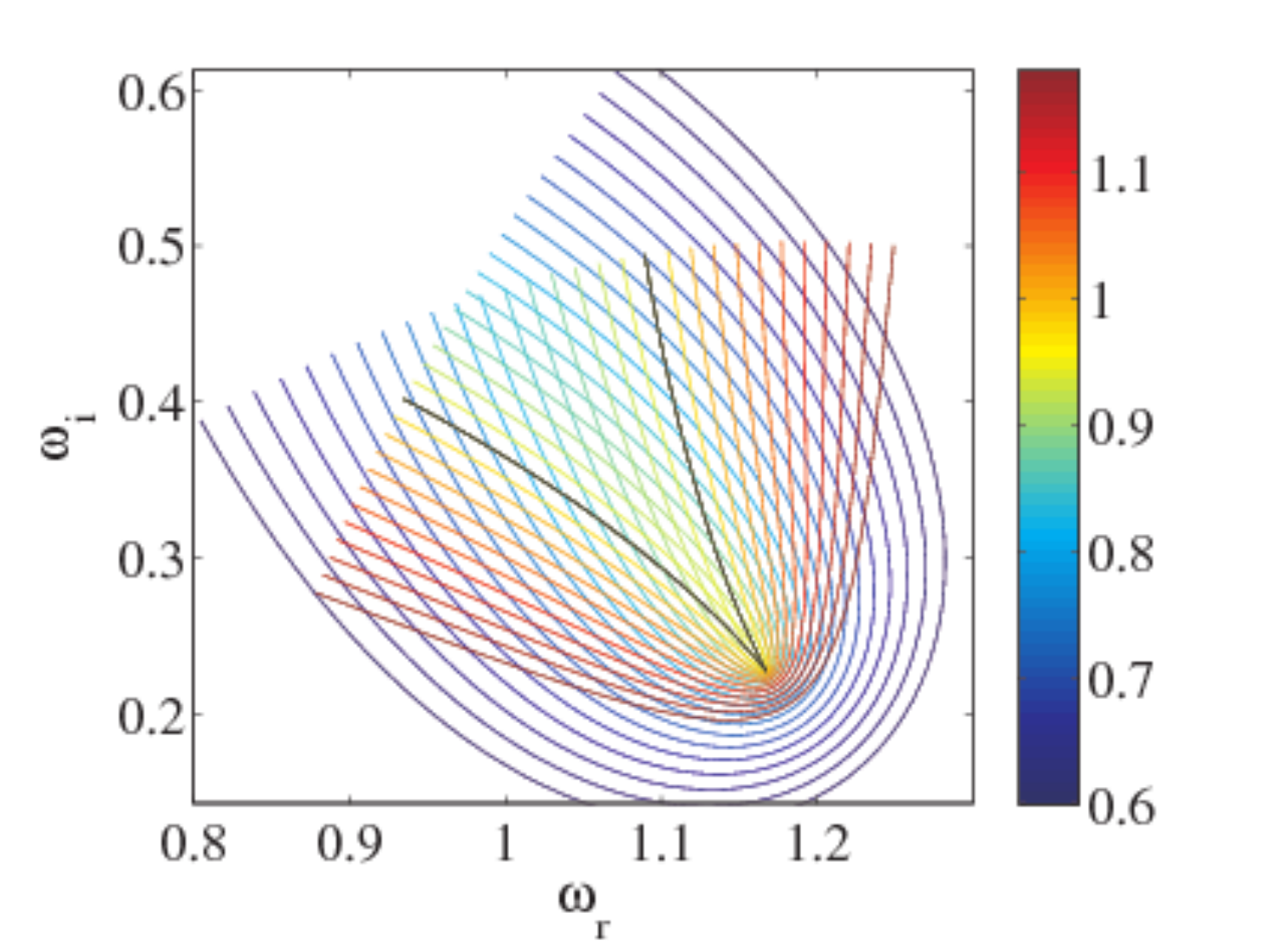}}
\caption{The global topography of $\omi$ for $(Re,\Lambda,N)=(100,-1.1,2)$.  Figure~(c) shows contours of $\ar$ in a complex $\omega$ plane.  The $\omega$ cusp corresponds to the $\alpha$-pinchpoint in Figures~(a) and (b).}
\label{fig:singular0}
\end{figure}

\begin{figure}[htb]
\centering
\includegraphics[width=0.45\textwidth]{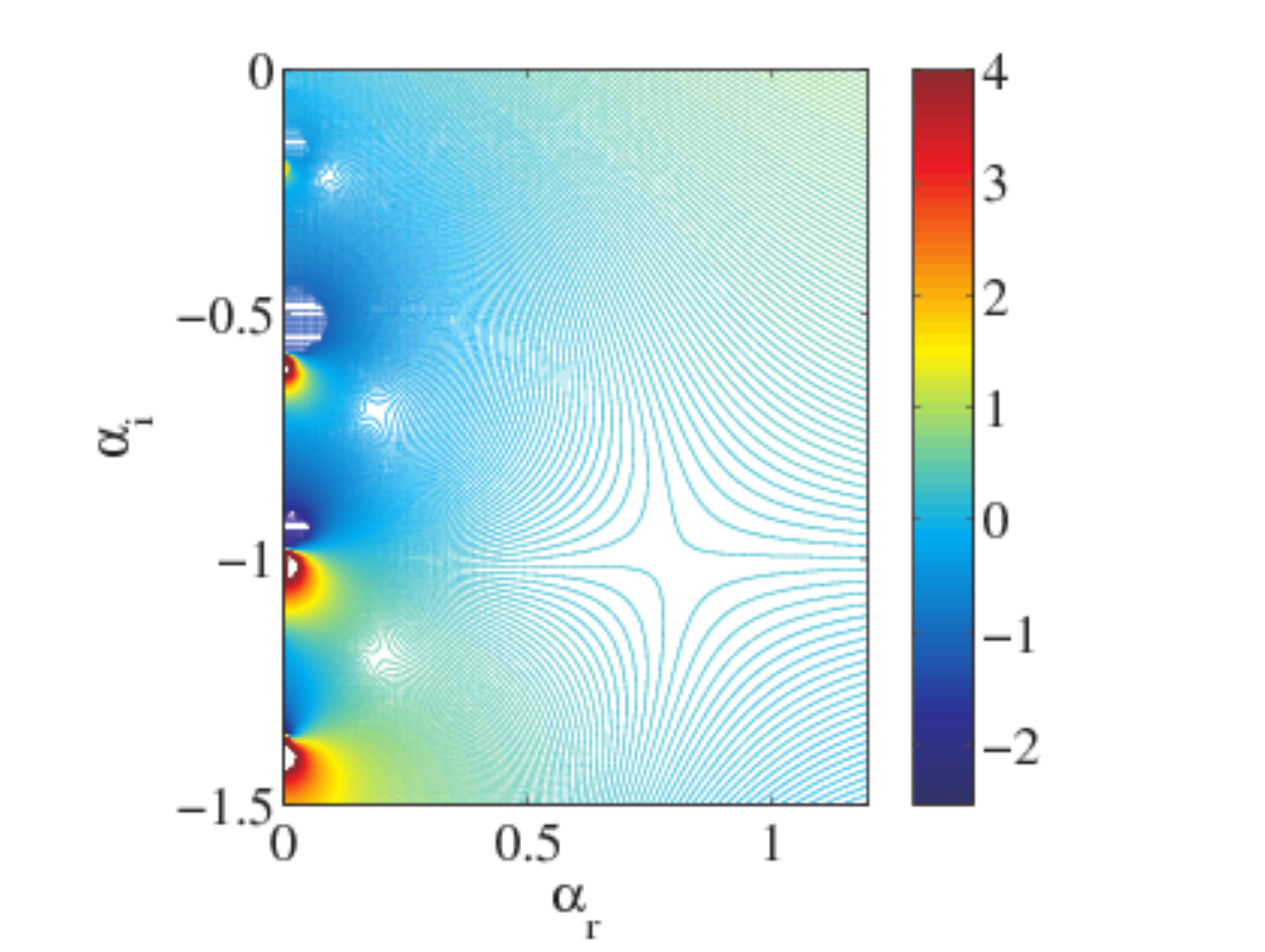}
\caption{The global topography of $\omi$ for $(Re,\Lambda,N)=(100,-1.1,5)$.}
\label{fig:singular1}
\end{figure}

Before applying Equation~\eqref{eq:omi_taylor}, we investigate the feasibility of so doing, that is, we examine the topography of $\omega(\alpha)$ and the proximity of any singularities in this function to the point of interest $(\ar,0)$ on the real line.  Each study is presented using numerical method (1), but further tests were carried out using  method (2), producing near-identical results.
A first study is carried  at the parameter values $(Re,\Lambda,N)=(100,-1.1,2)$.  Because only the sinuous mode produces absolute instability, the varicose mode is projected out of the problem using appropriate boundary conditions at the symmetry line $z=0$~\citep{Monkey1988}.   This removes a potential source of mode competition from the complex dispersion relation.  The results of these calculations are shown in Figure~\ref{fig:singular0}.
We first of all comment on the large peaks along the imaginary axis in Figure~\ref{fig:singular0}.  We have confirmed numerically that these are  poles (rather than large but finite values of $|\omi|$): successively decreasing the resolution in the scan of the complex $\alpha$-space leads to larger numerical maximum values of $(\omr,\omi)$.  Clearly, these poles are confinement poles (as in the works of~\citet{Juniper2006,Healey2007,Healey2009}).
We investigate the saddles induced by the confinement poles in  Figure~\ref{fig:singular0}(a).  Here,  the spatial branch corresponding both to the confinement saddle and the `physical' saddle are shown (the `physical' saddle is the one that persists in the limit as $H\rightarrow\infty$).  Only the physical saddle possesses spatial branches emanating from opposite half-planes and is therefore the only one that satisfies the Briggs criterion for contributing to absolute instability (the $\omega$-cusp at the $\alpha$-pinchpoint is shown in Figure~\ref{fig:singular0}(c)).  This result was verified further by a so-called \textit{ray analysis} (e.g. the work by~\citet{Onaraigh2012a}), where the growth-rate $\sigma(V)$ along rays moving with velocity $V$ with respect to the laboratory frame is computed from direct numerical simulation of the linearized equations of motion: again, $\sigma(V)$ possesses only one branch, associated with the physical saddle (the reader is referred to the Supplementary Material for details about such direct numerical simulations).  Furthermore, the path of steepest descent is  shown in Figure~\ref{fig:singular0}(b).  This path follows a line of constant $\omr$ and passes through the physical saddle of $\omi$ and points from the South West to the North East, and connects at infinity to the real axis. 
The dispersion relation for $(Re,\Lambda,N)=(100,-1.1,5)$ was found to contain a branch cut emanating from the point $(0.44,-1.79)$.  However, this occurred far below the saddle in the complex plane and was not collocated with the steepest-descent path.  Consideration of the growth rate $\sigma(V)$ computed via linearized DNS showed that this singularity did not contribute to the evolution of an initially-localized pulse.
 Finally,  dispersion relations for a further parameter set $(Re,\Lambda,N)=(100,-1.1,5)$ are shown in Figure~\ref{fig:singular1}.  These results are very similar to Figure~\ref{fig:singular0} (in particular, only the physical saddle is found to contribute to the absolute instability) and are not analysed in any further detail.

\begin{figure}[htb]
\centering
\subfigure[]{\includegraphics[width=0.48\textwidth]{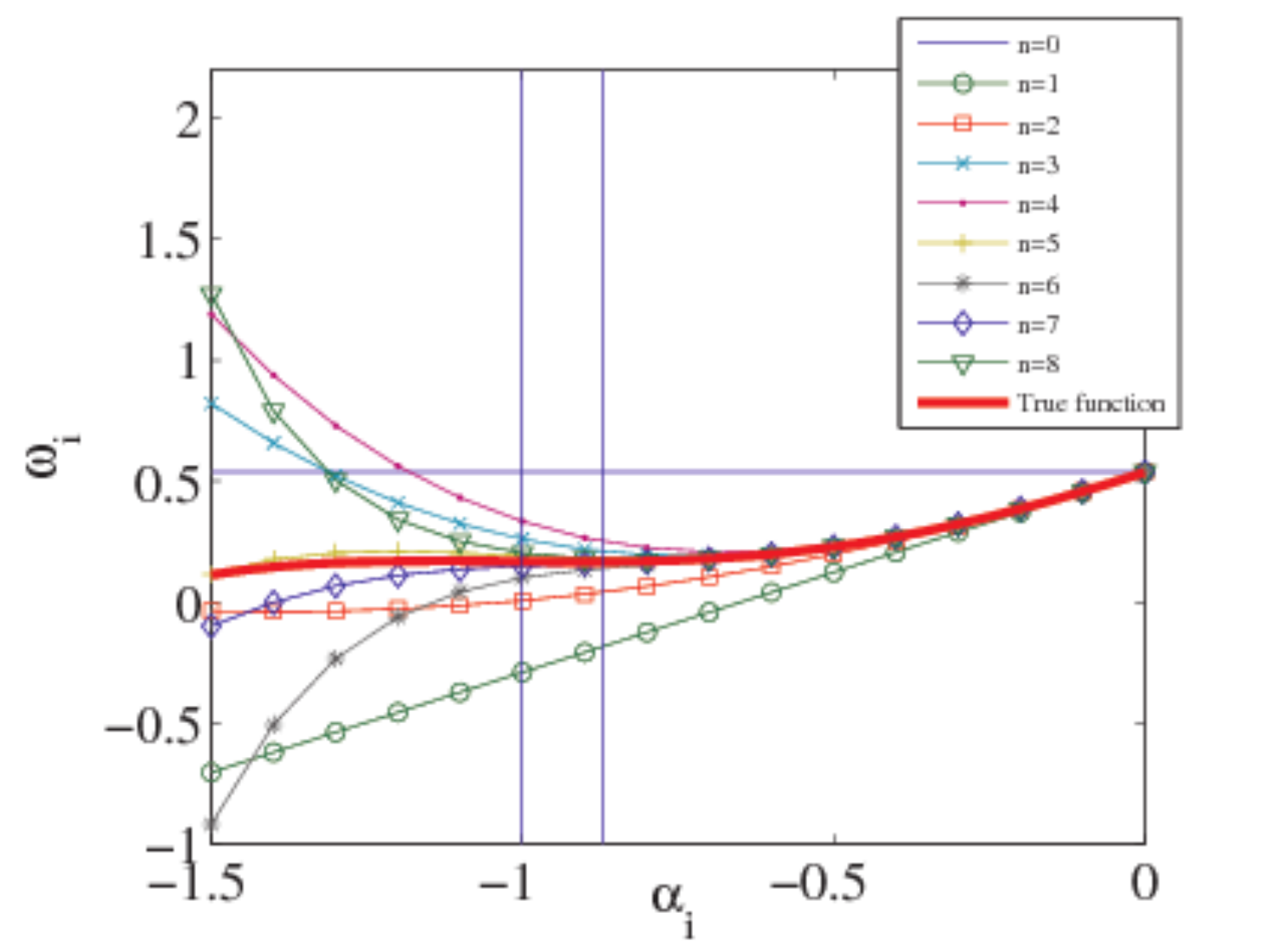}}
\subfigure[]{\includegraphics[width=0.48\textwidth]{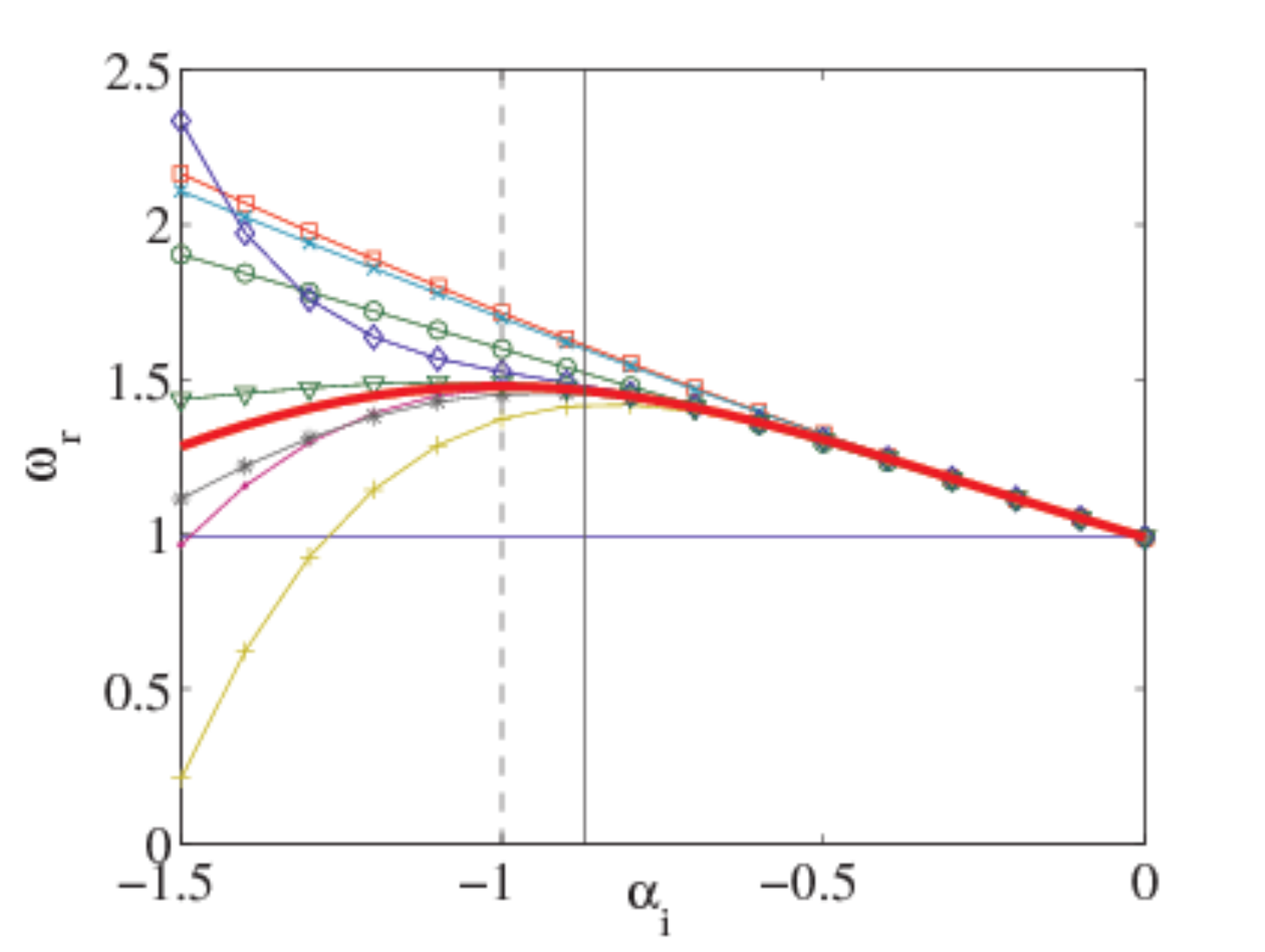}}
\caption{Successive truncations of (a) the series for $\omi$ (Equation~\eqref{eq:omi_taylor}); (b) an analogous series for $\omr$, and a comparison with the true functions for $(\omr,\omi)$.  Here $\ar=0.7729$ (i.e. a point close to the saddle),  $(Re,\Lambda,N)=(100,-1.1,2)$.  The solid vertical line indicates the radius of convergence, while the broken vertical line indicates the $\ai$-location of the saddle point.}
\label{fig:singular2}
\end{figure}

In Figures~\ref{fig:singular0}--\ref{fig:singular1} the singularity nearest to the real axis lies at $\ai=-0.18$ for both $N=2$ and $N=5$ (the dispersion relations are generated with confinement at $H=8$).
The physical saddle is located at $(\ar,\ai)=(0.94,-0.7)$ and $(0.79,-1.0)$ for $N=2$ and $N=5$ respectively (its computed location is the same for numerical methods (1) and (2) and is independent of $H$, for $H$ sufficiently large).
Thus, the radius of convergence defined by Equation~\eqref{eq:roc0} is  $R=0.96$ and $R=0.82$ for $N=2$ and $N=5$ respectively.  
It would therefore appear that the series~\eqref{eq:omi_taylor} converges for the $N=2$ case and diverges for the $N=5$ case; however, we have found that high-order truncations of the power series for $\omi$ (and an analogous series for $\omr$) 
give good approximations to $(\omr,\omi)$ outside of this region (Figure~\ref{fig:singular2}).  The numerical differentiation of the $\ar$-derivatives for Figure~\ref{fig:singular2} was been performed by interpolating the temporal dispersion relation on to a Chebyshev collocation grid and then performing analytic differentiation on the Chebyshev basis functions of this interpolation (this is a different Chebyshev collocation grid from the one discussed in \S\ref{sec:numerics}\ref{sec:numerics:method}).
In certain cases, a truncation of a divergent Taylor series can approximate a function reasonably well.  Indeed, for certain cases there exists an `optimal truncation', whereby some finite truncation order $N_{\mathrm{trunc,opt}}<\infty$ minimizes the difference between the generating function and the truncated Taylor series\footnote{For example, the function $f(x)=(1+x)\log (1+x)-x$.  On the interval $|x|<1$, this function can be represented as a power series.  However, for $x>1$ the associated series diverges, but an optimal truncation can be found that minimizes the difference between $f(x)$ and the truncated series.}.  Thus,  for physical applications such as the present one, it suffices to determine on a case-by-case basis whether finite truncations of the series for $\omi$ are good approximations to the underlying function.

Motivated by the good agreement between the true function $\omi(\ar,\ai)$ and finite truncations of its Taylor series in Figure~\ref{fig:singular2}, we have investigated the purely temporal stability properties of the system in Figure~\ref{fig:singular2} with a view to applying Equation~\eqref{eq:omi_taylor} to the spatiotemporal problem.
\begin{figure}[htb]
\centering
\subfigure[]{\includegraphics[width=0.4\textwidth]{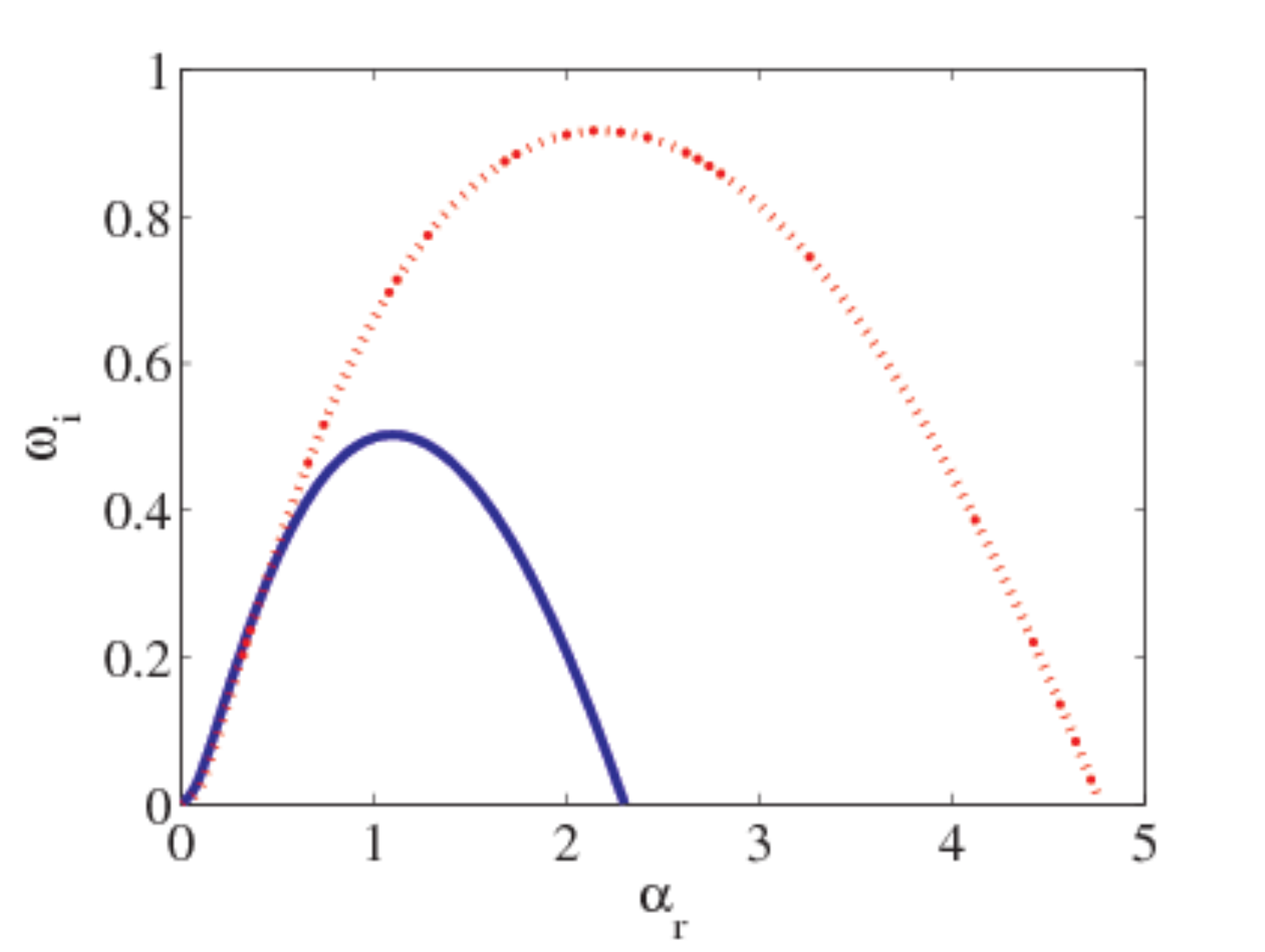}}
\subfigure[]{\includegraphics[width=0.4\textwidth]{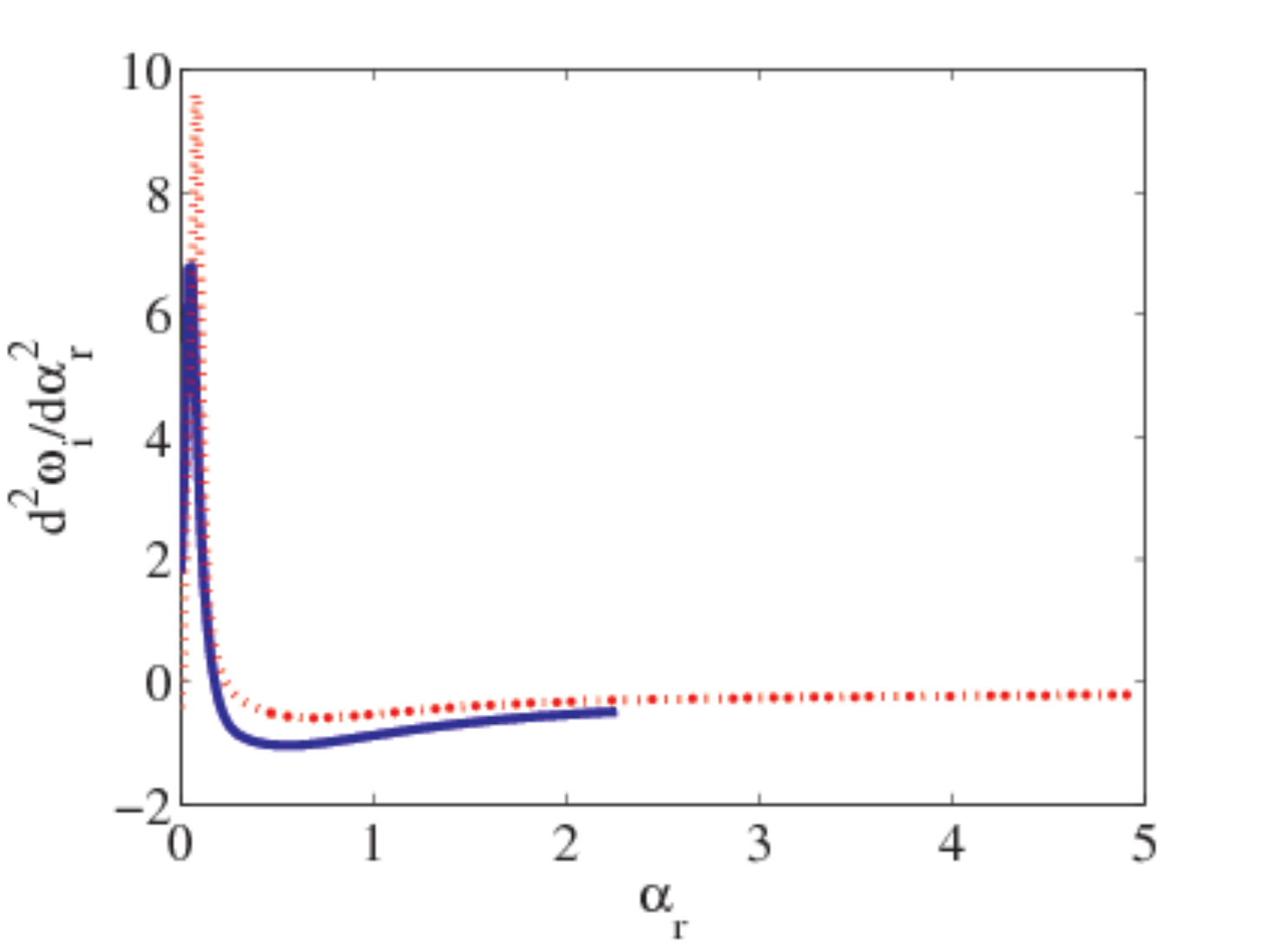}}\\
\subfigure[]{\includegraphics[width=0.4\textwidth]{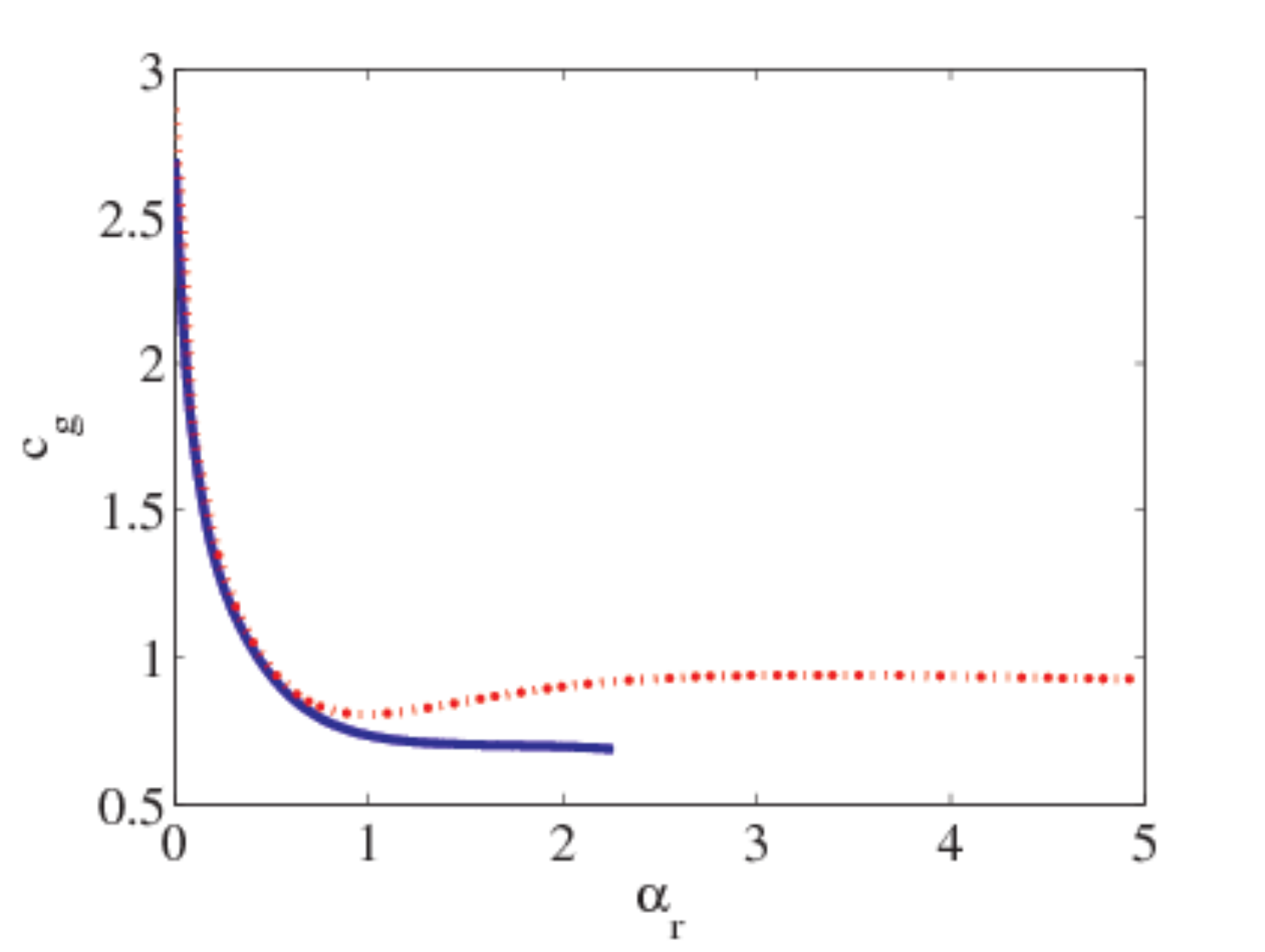}}
\subfigure[]{\includegraphics[width=0.4\textwidth]{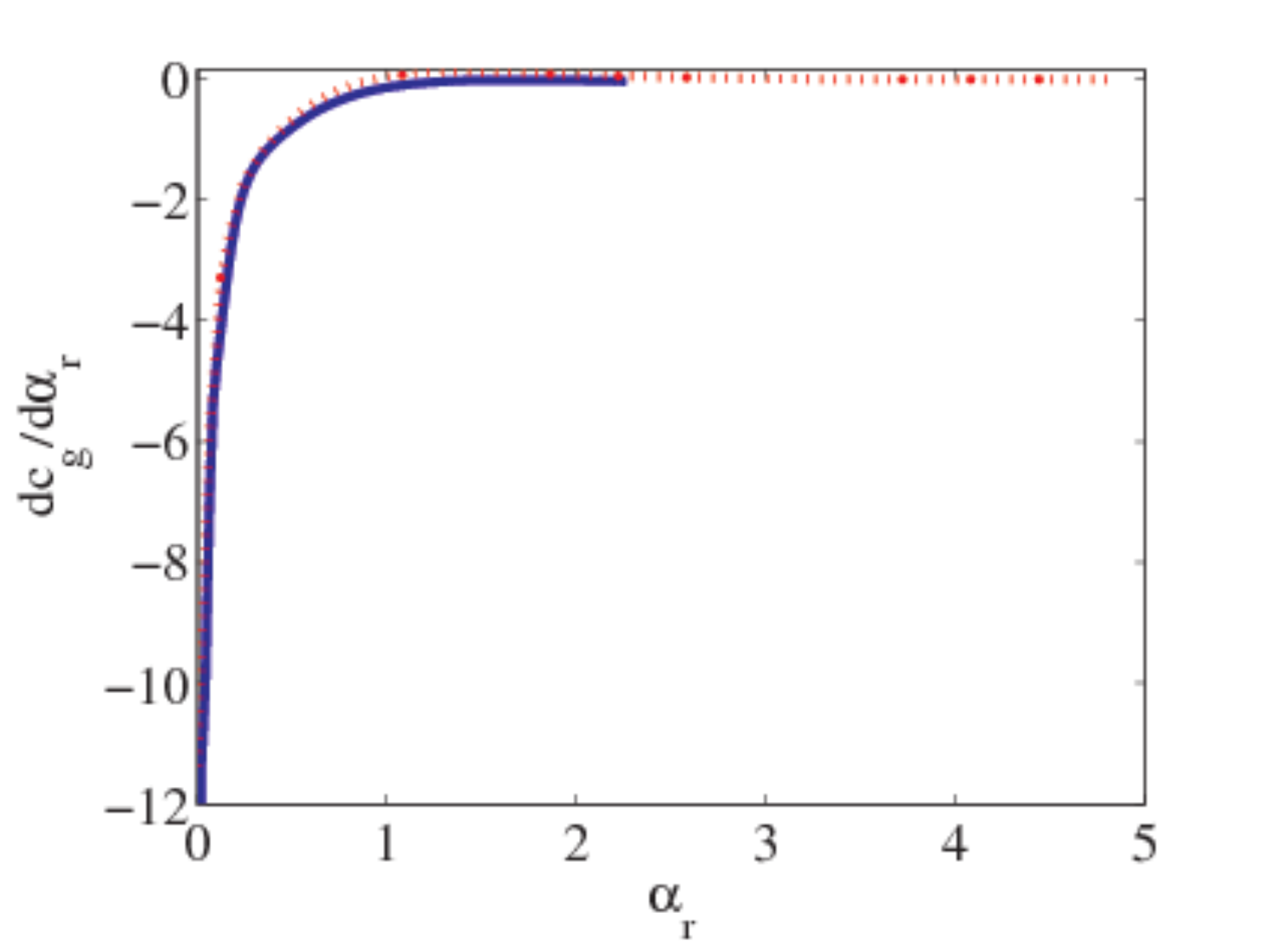}}
\caption{Dispersion curves for two test cases with $(Re,\Lambda)=(100,-1.1)$.  Solid line: $N=2$; broken line: $N=5$.  The curves for the $N=2$ case are not continued beyond $\ar\approx 2.3$ because of the sharp mode competition that occurs there.  However, this is a region where all modes are stable and is therefore not relevant to the dynamics.  }
\label{fig:bluff1}
\end{figure}
This is done in Figure~\ref{fig:bluff1}: the growth rate has the basic features of a quadratic equation in $\ar$, while the group velocity depends only weakly on the wavenumber $\ar$.  However, the second and first derivatives of these quantities are non-linear and non-constant respectively,
and the assumptions of the  quadratic approximation do not apply.  We therefore examine both the quadratic approximation and further, higher-order approximations; the latter involve a truncation of Equation~\eqref{eq:omi_taylor} at cubic (or higher) order, together with a procedure analogous to Equations~\eqref{eq:omi_approx}--\eqref{eq:saddle_sign}  to obtain the criterion for the onset of absolute instability. 
%
%
%
%

The true boundary between the convective and absolute regions is calculated numerically from a spatio-temporal OS analysis and is identified according to steps (i)--(ii) in \S\ref{sec:intro}.  
\begin{figure}[htb]
\centering
\subfigure[]{\includegraphics[width=0.4\textwidth]{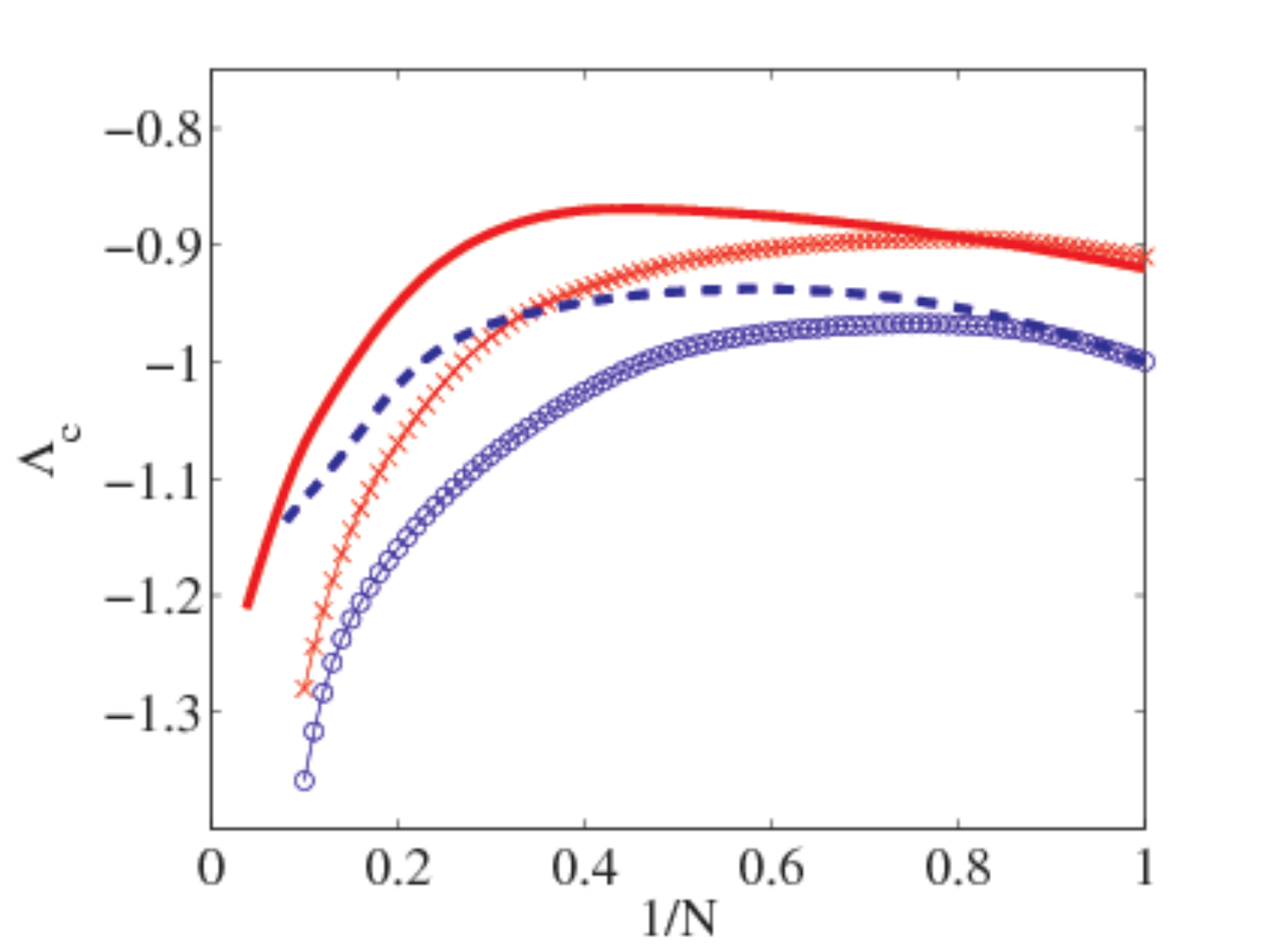}}
\subfigure[]{\includegraphics[width=0.4\textwidth]{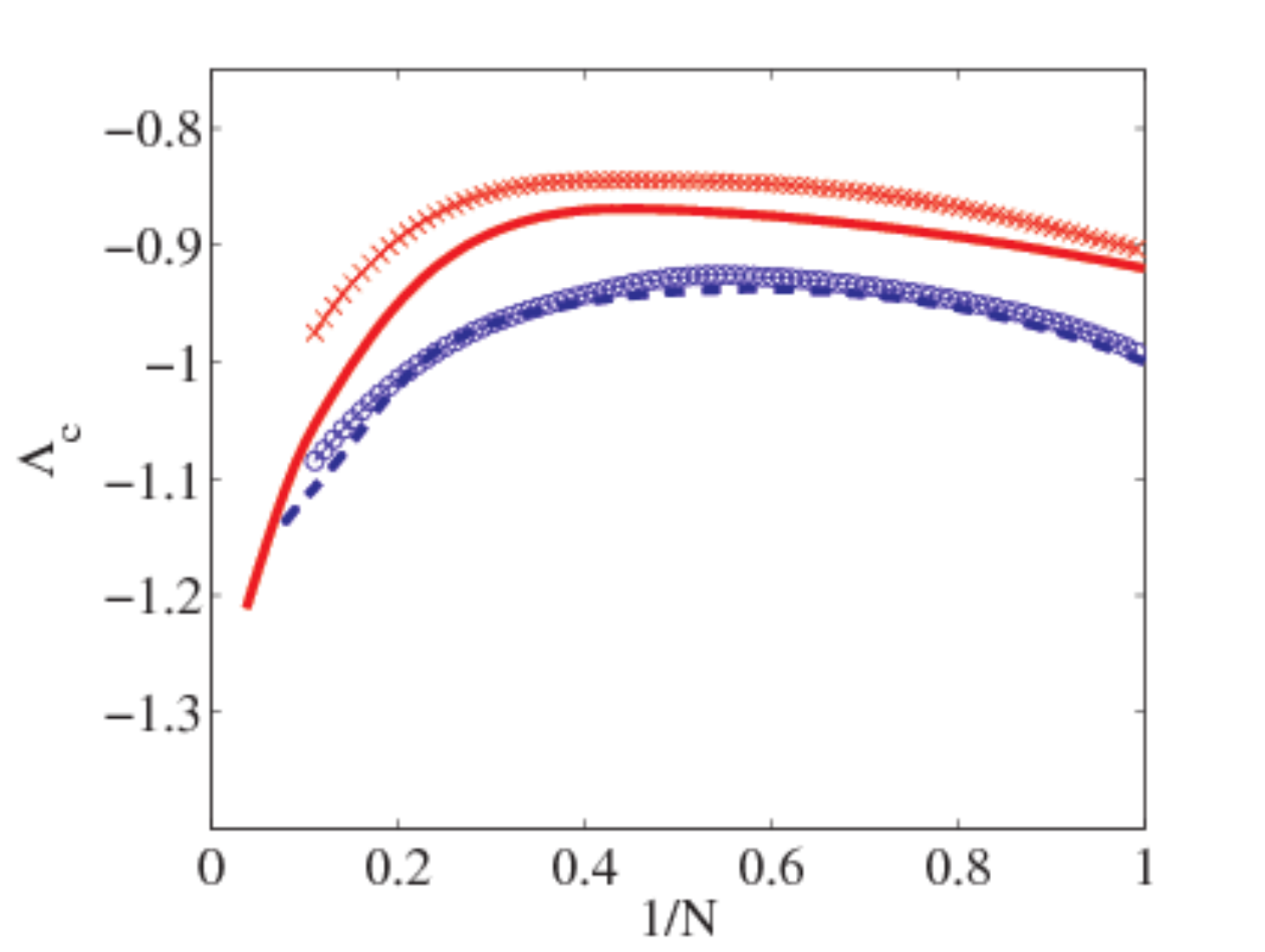}}
\caption{Boundary between the absolute and convective regions for different Reynolds numbers, with comparison to the (a) quadratic, and (b) cubic approximations.  Curves without symbols correspond to direct OS eigenvalue calculations in the complex $\alpha$-plane and agree exactly with the results by~\citet{Monkey1988}.  
 Dashed curve: $Re=20$; solid curve: $Re=100$.  Circles: approximate curves at $Re=20$; crosses: approximate curves $Re=100$.  
}
\label{fig:bluff2}
\end{figure}
The estimated boundary is calculated, according to quadratic and cubic approximations.   The quadratic approximation yields a  maximum error of $20\%$ with respect to the numerically-generated curves.  For the cubic approximation, the $Re=20$ approximate curve and the numerically-generated curve virtually coincide, while the maximum error in the $Re=100$ curve for the cubic approximation is reduced to  $7.5\%$ (Figure~\ref{fig:bluff2}).  We have verified that going to higher-order truncations in Equation~\eqref{eq:omi_taylor} for the $Re=100$ case leads to near-perfect overlap between these curves  (a maximum error of $2.5\%$ in a fifth-order truncation, at $Re=100$). 
%
%
%

\subsection{Interpretation of results via the quadratic approximation}

\begin{figure}[htb]
\centering
\subfigure[]{\includegraphics[width=0.49\textwidth]{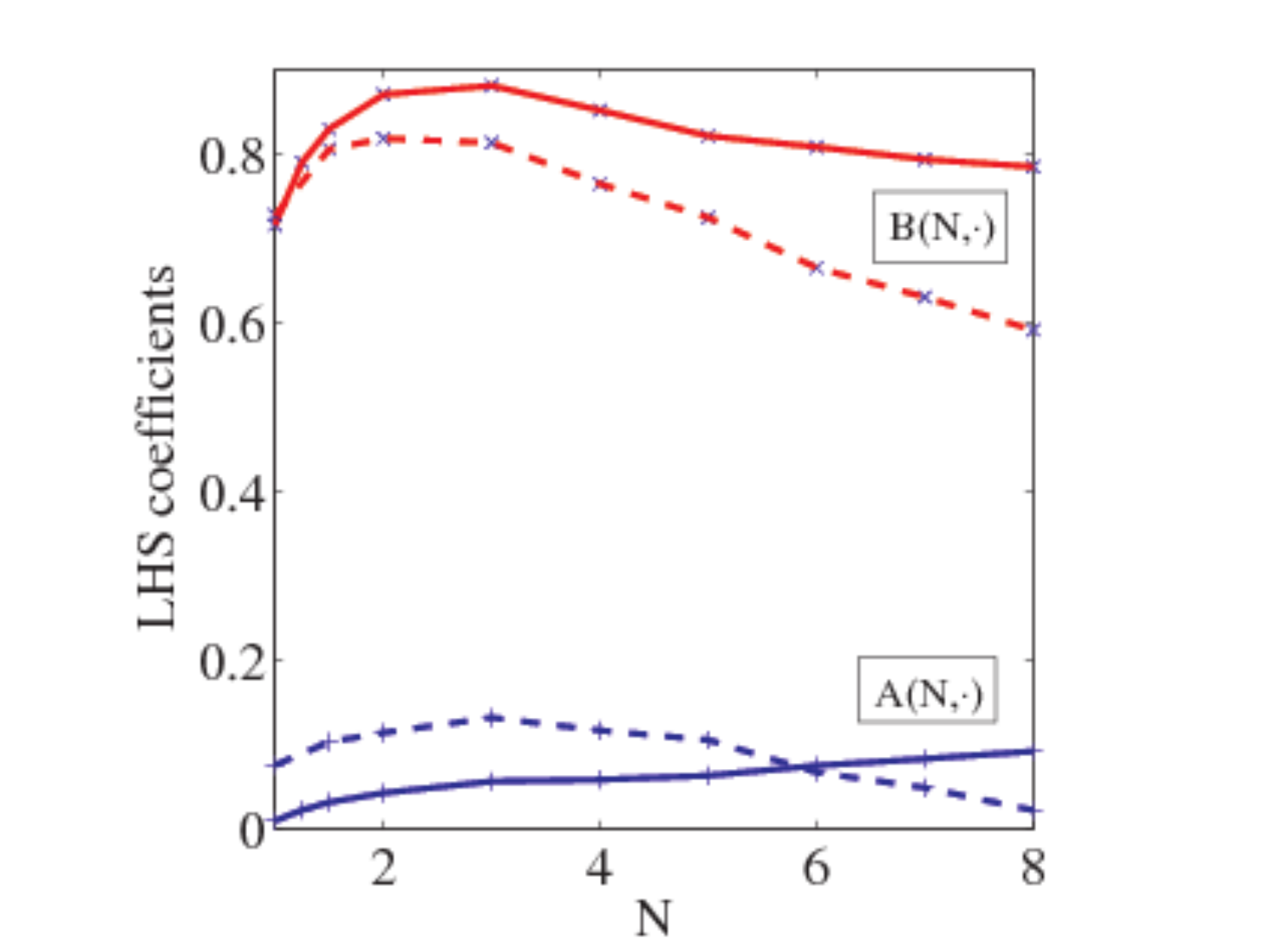}}
\subfigure[]{\includegraphics[width=0.49\textwidth]{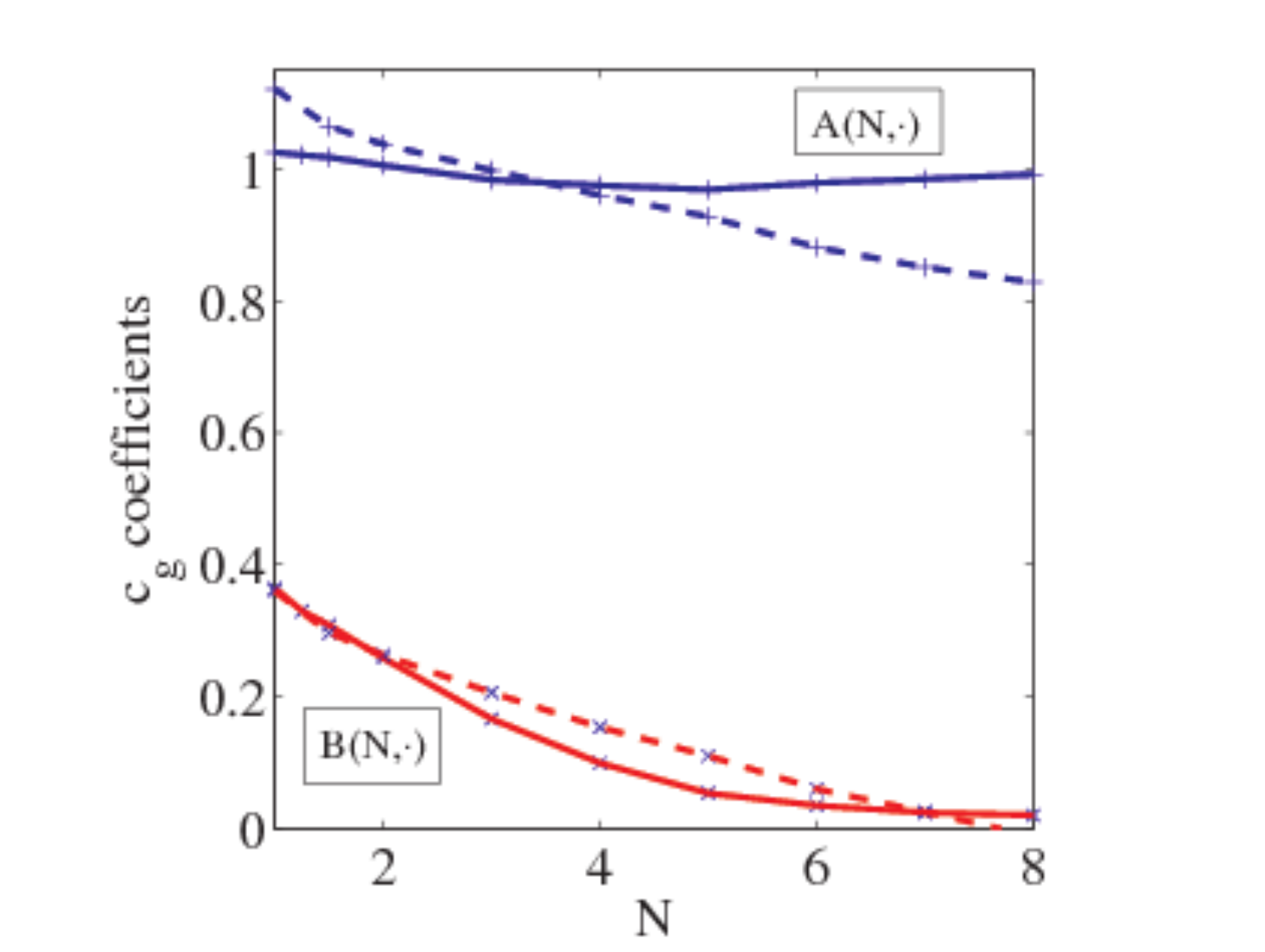}}
\caption{Dependency on $N$ of coefficients in fits linear in $\Lambda$ of the left-hand side of Equation~\eqref{eq:saddle_sign} (a) and the group velocity (b) at $Re=100$ (solid lines) and $Re=20$ (dashed lines); plusses and crosses represent $A$ and $B$, respectively. In (a), the sign of both coefficients has been reversed.
}
\label{fig:lhscg}
\end{figure}

The quadratic approximation can now be used to further analyse the C/A transition.  The coefficients in a quadratic approximation for the temporal dispersion relation $\omitemp(\ar)$ can be related to the temporally most-dangerous mode $(\am,\omim)$, such that
\begin{equation}
\omitemp(\ar)=2\omim\left[\frac{\ar}{\am}-\tfrac{1}{2}\left(\frac{\ar}{\am}\right)^2\right],\qquad\text{quadratic approximation}.
\end{equation}  
Moreover, we have found in our numerical calculations that the $\ar$-location of the unstable saddle point in the complex dispersion relation almost coincides with the most-dangerous mode, such that (using the notation in \S\ref{sec:limiting}\ref{sec:limiting:quadratic}) $\ar^*\approx \am$.  The criterion~\eqref{eq:saddle_sign} in \S\ref{sec:limiting}(b) describing the onset of absolute instability  then reduces to
\begin{equation}
{2\omim}\slash{\am}=\cg (\am).
\label{eq:CA_QA}
\end{equation}
Both sides of the condition~\eqref{eq:CA_QA} are well represented by expressions of the form $A(N,Re)+B(N,Re)\Lambda$ over the part of parameter space considered in \S\ref{sec:numerics}\ref{sec:numerics:results}. In Figure~\ref{fig:lhscg} the  dependencies of these 
coefficients on $N$ and $Re$ are shown. The most significant of these is the $N-$dependency of the coefficient $B$ for the group velocity: increasing $N$ makes the group velocity approach unity rapidly from below. If we ignore for the time being the other dependencies on $N$ and use for the left-hand side of Equation~\eqref{eq:CA_QA} simple $N-$averaged values ($A\equiv a_0\approx -0.05$ and $B\equiv a_1\approx -0.8$ for $Re=100$), and for the group velocity the curve fit
\begin{equation}
c_g ( \alpha_m ) \approx  1+b_0{\rm exp}(-b_1N)\Lambda
\label{eq:cgCA}
\end{equation}
(with $b_0\approx 0.6$ and $b_1\approx 0.5$ for $Re=100$), we obtain an approximate expression for the C/A transition based on the quadratic approximation:
\begin{equation}
\Lambda_{CA}\approx  \frac{a_0-1}{b_0{\rm exp}(-b_1N)-a_1}.
\label{eq:appCA}
\end{equation}
We have found this to represent the full results obtained with the quadratic approximation within the uncertainty of the value of the coefficients.  Overall then, the main trend is that in making $\Lambda$ more negative, the group velocity (the right-hand side of Equation~\eqref{eq:CA_QA}) decreases, whereas the left-hand side of Equation~\eqref{eq:CA_QA} increases; these two effects promote  absolute instead of  convective instability. Thus, the steep drop in the curves at large $N$ in Figure~\ref{fig:bluff2} is associated with the fact that increasing $N$  increases the group velocity.
\begin{figure}[htb]
\centering
\subfigure[]{\includegraphics[width=0.49\textwidth]{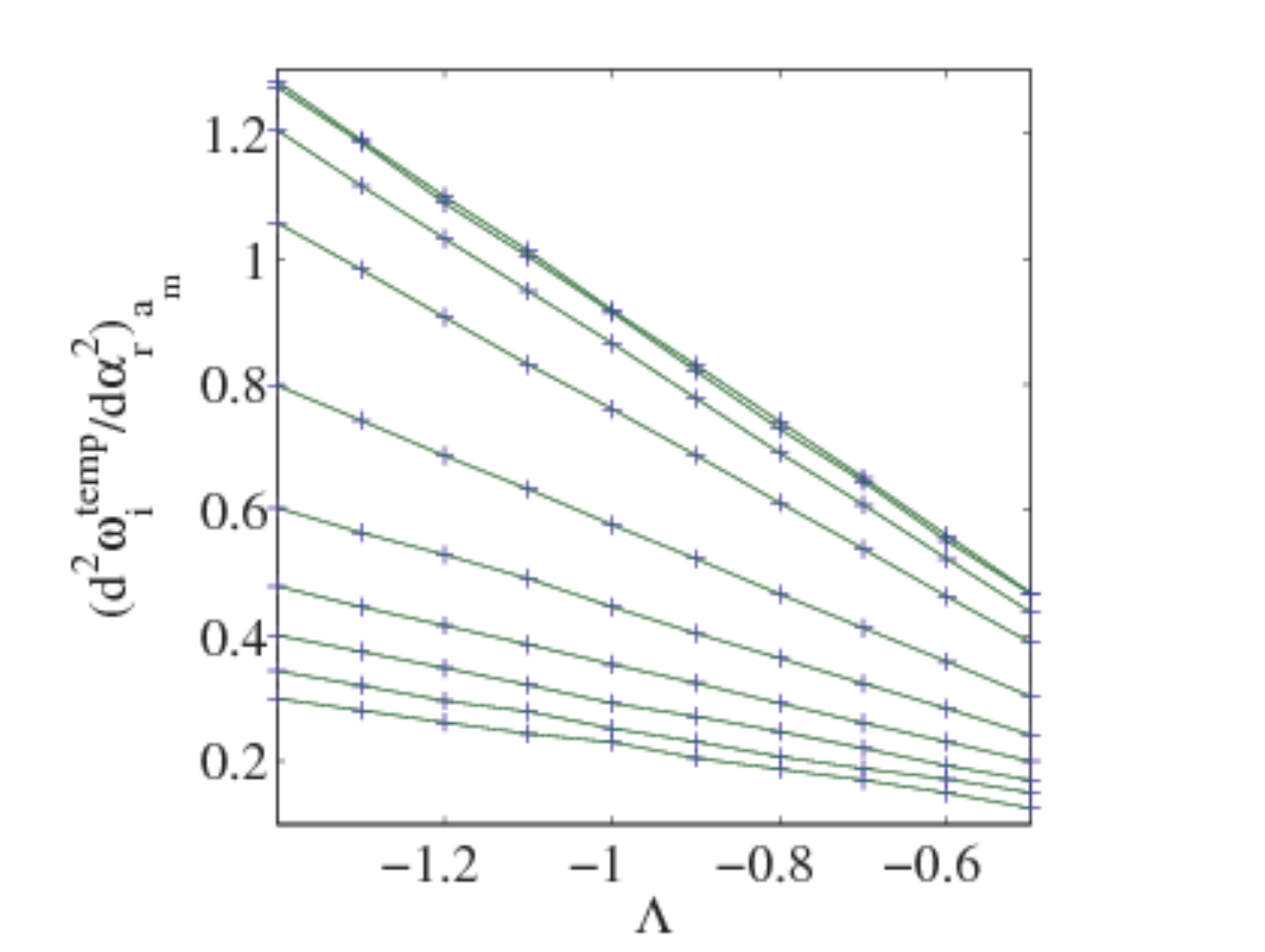}}
\subfigure[]{\includegraphics[width=0.49\textwidth]{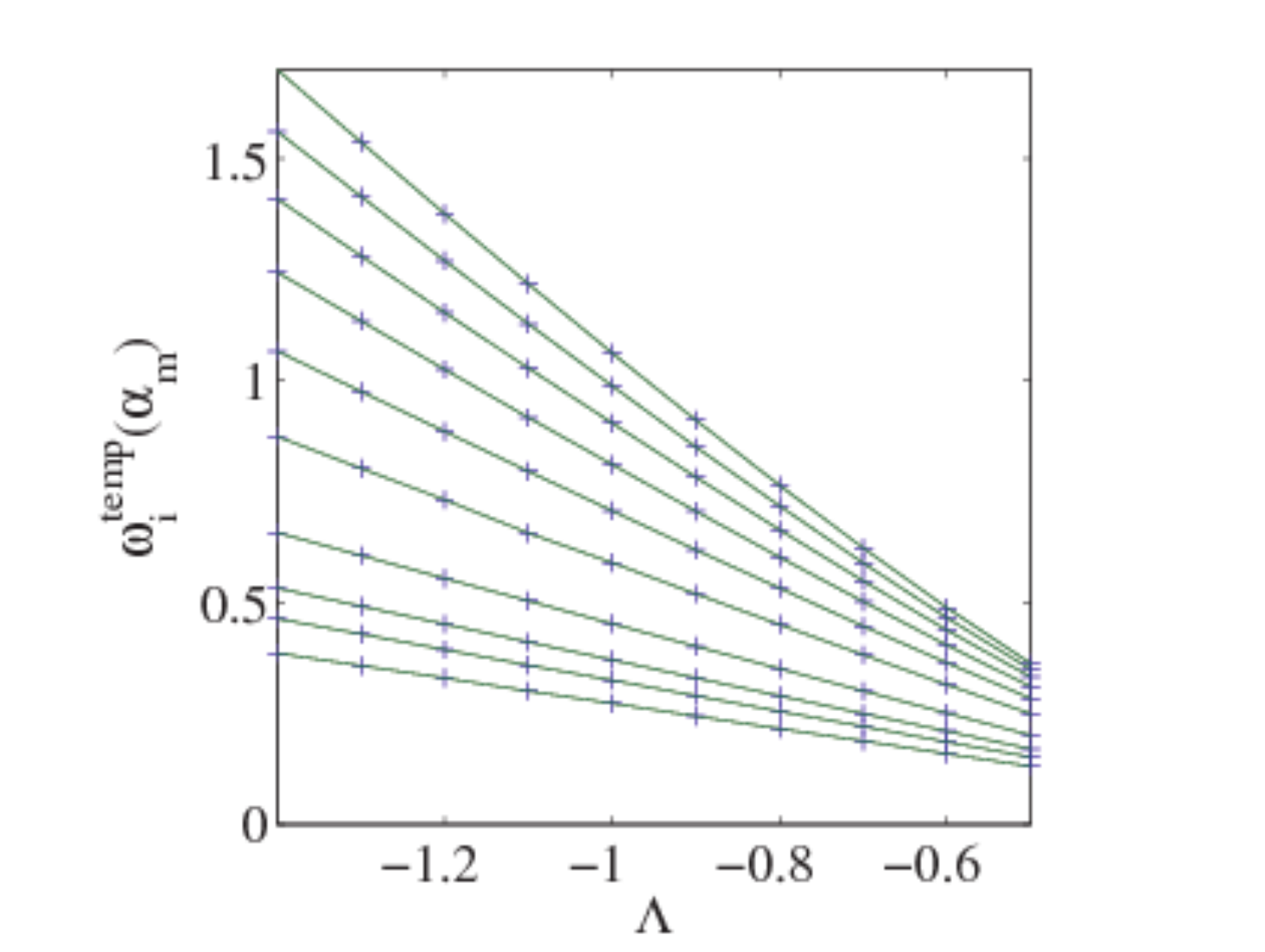}}
\caption{The dependencies on $N$ and $\Lambda$ of (a) $-d^2\omitemp/d\ar^2$; (b) $\omitemp$, at the most-dangerous temporal mode for $Re=100$. Results are shown for $N=1,1.25,1.5,2,3,4,5,6,7,8$ from top to bottom in (a), and from bottom to top in (b).
}
\label{fig:om_ddom}
\end{figure}

 Regarding the dependencies on $Re$, we first note that at low values of $N$, the magnitude of the left-hand side of Equation~\eqref{eq:CA_QA} is small at $Re=20$ compared with $Re=100$, whereas the magnitude of the right-hand side is large; these changes both favour convective instability. At intermediate-to-large values of $N$, both terms decrease when the value of $Re$ is decreased, with the left-hand side more so than the right-hand side.  Hence, the C/A transition in Figure~\ref{fig:bluff2} occurs at a more negative value of $\Lambda$ for the lowest value of $Re$ due to effects on both sides in Equation~\eqref{eq:CA_QA}.  These effects can be accounted for in the above through the introduction of explicit $Re-$dependencies in the coefficients $a_0,a_1,b_0,b_1$ (even if only representing the $Re-$dependencies of their $N-$averaged values).   These dependencies are already visible in Figure~\ref{fig:lhscg}, although their parametrization   is beyond the scope of this paper.

These somewhat crude approximations can be refined by representing the dependency of the left-hand-side coefficients $A$ and $B$ on $N$, the causes of which we first briefly investigate here. At values of $N$ near unity, the magnitude of the left-hand side of Equation~\eqref{eq:CA_QA} first increases strongly with $N$ before decreasing beyond $N\approx 3$, such that the region of absolute instability in parameter space is reduced further at large values of $N$. In fact, the results in Figure~\ref{fig:lhscg}(a) tell two different stories, for values below and above $N=3$. Referring back to Equation~\eqref{eq:saddle_sign}, we recall that the left-hand side is the product of the maximum temporal growth rate and minus the curvature of the dispersion relation. Results for these two components are shown in Figure~\ref{fig:om_ddom}. We conclude from  Figure~\ref{fig:om_ddom} that $\omitemp(\am)$  increases uniformly with increasing $N$, consistent with the increase in the entire left-hand side of  Equation~\eqref{eq:saddle_sign} at low values of $N$, and that the resulting promotion of absolute instability is only reversed by a decrease in $-d^2\omitemp/d\ar^2$ at large values of $N$; at low values of $N$, this curvature term hardly depends on $N$. It is possible to account for these trends in  Equation~\eqref{eq:appCA}, by using bilinear fits for $\omim$ and $\am$ (as well as Equation~\eqref{eq:cgCA} for the group velocity):
%
%
\begin{equation}
\am  \approx a_\alpha^0+a_\alpha^1N+b_\alpha^0\Lambda+b_\alpha^1N\Lambda,\qquad
\omim \approx  a_\omega^0+a_\omega^1N+b_\omega^0\Lambda+b_\omega^1N\Lambda.
\label{eq:CA_morefits}
\end{equation}
Substitution of Equation~\eqref{eq:CA_morefits} into Equation~\eqref{eq:CA_QA} then yields a quadratic equation for $\Lambda$ at the C/A transition.
This whole approach can be extended by lifting the approximation $\ar^*\approx \am$, or by going to higher-order truncations of the dispersion relation; however, these complications obscure the  understanding achieved here using simple linear and bilinear fitting, and such extensions are not pursued further here.

\section{Discussion}
\label{sec:disc}

We have presented an analytical connection between spatio-temporal and temporal growth rates in a local linear stability analysis. Specifically, we have derived  criteria for a transition between convective and absolute instability, explicitly in terms of the temporal problem, to increasingly refined levels of approximation.  The simplest approximation is a quadratic one where the location of the saddle $\ar^*$ is assumed to coincide with the temporally most-dangerous mode; then, the criterion for the onset of absolute instability is given by Equation~\eqref{eq:CA_QA}.   At the next level of approximation (still quadratic), one determines  $\ar^*$  via Equation~\eqref{eq:saddle}.
 Although this approach can be used directly to determine C/A transitions, we anticipate that the main use of the theory will be in analysing results obtained with an efficient computational algorithm for the full OS problem. 
We imagine that one would perform such a numerical calculation, diagnose absolute instability (using the saddle-point/pinch-point criteria), and then characterize the instability in detail. Our formula can help in this characterization in the following ways: 
%
%

\textit{1.}  In parametric studies, the dependency of C/A transition curves on model parameters can be difficult to analyse, especially if there are many governing parameters. In \S\ref{sec:numerics} we have seen an example wherein  simple approximations based on Equation~\eqref{eq:omi_taylor},  together with insight into the trends exhibited by the dispersion relation in the purely temporal problem, yield a detailed explanation of the behaviour of the C/A transition curves.


\textit{2.}  Temporal mode competition is observed in various systems and, as pointed out in the Introduction, disentangling different spatio-temporal modes poses a formidable problem. The present approach then provides a convenient aid: for each temporal branch, an approximation of the corresponding spatio-temporal growth rate in the complex $\alpha$ plane results directly from  Equation~\eqref{eq:omi_taylor}. An example is given in Figure~\ref{fig:singular2}.  Although only a cross-section for a single value of $\ar$ is shown in Figure~\ref{fig:singular2}, this is readily extended to entire sections of the complex $\alpha$ plane. Indeed, in computing the C/A transition in truncations  higher than quadratic, we have found this approach more straightforward rather than the root-finding approach taken for the relatively simple quadratic case (\S\ref{sec:limiting}).

\textit{3.}  It follows from the previous point that features in contour plots for a spatio-temporal growth rate (such as saddle points) can be related directly to a specific temporal mode with the present approach. The advantage of this is that such features can then be associated with the physical mechanism that governs that temporal mode.
%
%
%
The application of the approach proposed here to real flows has recently been completed in~\citet{Onaraigh2012a}, where we have studied absolute and convective instability in two-phase gas/liquid flows.  There, however, the focus was on the application of the theory in a detailed parameter study and a derivation of the quadratic approximation was not included; in the present work, the focus is on the mathematical derivation of Equation~\eqref{eq:omi_taylor}, discussion of the radius of convergence of the formula, and the investigation of the accuracy of finite truncations of the equation~\eqref{eq:omi_taylor}.  Such analysis is important before the theory can be used with confidence in further applications.


Throughout this paper we have considered two-dimensional disturbances, taking $\omega$ as a complex-valued function of a single complex variable (the complex wavenumber). However, this approach could be extended to three-dimensional spatio-temporal disturbances, via a double-Taylor-series expansion of the dispersion relation in the streamwise and spanwise wavenumbers. For cases wherein the spanwise wavenumber is real, the analysis presented herein requires no modification, as the spanwise wavenumber would then only enter as an additional parameter.

We conclude by discussing briefly a further application of this method to weakly globally unstable modes.  In nonparallel flows, the properties of the system vary in the streamwise direction, and a standard normal-mode decomposition is no longer possible.  The generic linear-stability problem to be solved then reads $\partial_t\psi=\mathcal{L}(\partial_x,\partial_z)\psi$, 
%
%
where $\mathcal{L}$ is a linear operator with coefficients that depend on the streamwise direction; the normal-mode decomposition $\partial/\partial x \rightarrow \imag \alpha$ is no longer possible.  We make trial substitution $\psi\rightarrow \psi(x,z)\mathe^{-\imag \omega_G t}$ and obtain the global eigenvalue problem, $\-\imag \omega_G\psi(x,z)=\mathcal{L}(\partial_x,\partial_z)\psi(x,z)$.
%
%
%
%
%
The system is globally unstable if there exists a global mode for which $\Im(\omega_G)>0$.  We introduce a parameter $\smallparam=\lambda/L$, where $\lambda$ is the typical length scale of the disturbance $\psi$, and $L$ is a measure of the system's extent in the streamwise direction. If $\smallparam\ll 1$, and if the coefficients  of the linear operator $\mathcal{L}$ depend only weakly on space, through the combination $X:=\smallparam x$, then a WKB approximation~\citep{huerre90a,Chomaz1991,Schmid2001} can be used to demonstrate that the leading-order approximation to the most-dangerous global mode is $\omega_G= \omega_0(X_\mathrm{s})+O(\smallparam)$,
%
%
where $\omega_0(X_\mathrm{s})$ is the saddle-point frequency calculated in the usual manner from a purely \textit{local} stability analysis, at the fixed parameter value $X_\mathrm{s}$, and $X_\mathrm{s}$ is the location of the saddle point of the complex-valued function $\omega_0(X)$ in the complex $X$-plane: $(\partial\omega_0/\partial X)(X_\mathrm{s})=0$.
%
%
The flow is therefore globally unstable if $\omega_0(X_\mathrm{s})>0$.
Thus, we are interested in a prescribed complex-valued function of a complex variable, $\omega_0(X)$, and its saddle point(s).  If $\omega_0(X)$ is holomorphic, and if $\omega_0(X)$ is known along the real line, then the arguments of \S\ref{sec:exact} apply also to the function $\omega_0(X)$, and the following formula pertains:
\begin{multline}
\omega_{0\mathrm{i}}(\xr,\xxi)=\omega_{0\mathrm{i}}(\xr,\xxi=0)+\sum_{n=0}^\infty \frac{(-1)^n}{(2n+1)!}\frac{d^{2n+1}}{d\xr^{2n+1}}\left[\omega_{0\mathrm{r}}(\xr,\xxi=0)\right]\xxi^{2n+1}\\
+\sum_{n=0}^\infty \frac{(-1)^{n+1}}{(2n+2)!}\frac{d^{2n+2}}{d\xr^{2n+2}}\left[\omega_{0\mathrm{i}}(\xr,\xxi=0)\right]\xxi^{2n+2}.
\label{eq:omi_taylor_global}
\end{multline}
Furthermore, if $\omega_0(X)$ is quadratic in $X$, then the truncation in \S\ref{sec:limiting}\ref{sec:limiting:quadratic} applies, and the condition for the onset of global instability reads
\begin{equation}
-\left[\frac{d^2\omega_{0\mathrm{i}}(\xr,\xxi=0)}{d\xr^2}\right]_{\xr^*}\omega_{0\mathrm{i}}(\xr^*,\xxi=0)=\tfrac{1}{2}\left[\frac{d\omega_{0\mathrm{r}}(\xr,\xxi=0)}{d\xr}\right]^2_{\xr^*},
\label{eq:saddle_sign_global}
\end{equation}
where $\xr^*$ satisfies a root-finding condition analogous to Equation~\eqref{eq:saddle}.  
%
A first-order truncation Equation~\eqref{eq:omi_taylor_global} was used in \cite{Hammond1997}, with $\omega_{0\mathrm{i}}(\xr,\xxi=0)$ determined numerically. The present approach would not only allow for rapid extension to higher order, but also open up the possibility of using Equation~\eqref{eq:saddle_sign_global} in conjunction with for instance the local quadratic approximation~\eqref{eq:saddle_sign} to determine the terms that make up Equation~\eqref{eq:saddle_sign_global}.

\subsection*{Acknowledgements}

This research was supported by the Ulysses-Ireland/France Research Visits Scheme, a programme for research visits between Ireland and France, jointly funded and administered by The Irish Research Council, the Irish Research Council for Science Engineering and Technology and Egide, the French agency for international mobility, with participation from the French Embassy in Ireland  and Teagasc.

L. \'O~N. would like to thank  R. Smith and C. Boyd for helpful discussion, and for a further helpful and interesting discussion with J. Healey. P.~S. would like to acknowledge useful discussions with his colleague B. Pier.

\bibliographystyle{plainnat}

\end{document}


%
%
%
%
%

\renewcommand\Authfont{\scshape}
\renewcommand\Affilfont{\small\normalfont}

\newcommand*{\TitleFont}{%
      \usefont{\encodingdefault}{cmr}{b}{n}
      \fontsize{16}{20}%
      \selectfont}

\sectionfont{\large}

\title
{\TitleFont An analytical connection between temporal and spatio-temporal growth rates in linear stability analysis -- Supplementary material}
\author[1]{\MakeUppercase Lennon \'O N\'araigh}
\author[2]{\MakeUppercase Peter D. M. Spelt}
\affil[1]{School of Mathematical Sciences, University College Dublin, Belfield, Dublin 4}
\affil[2]{ Laboratoire de M\'{e}canique des Fluides \& 
d'Acoustique, CNRS, Ecole Centrale Lyon, Ecully, France, and D\'{e}partement de M\'{e}canique, Universit\'{e} de Lyon 1, France}
\date{\small \today}
\maketitle

\begin{abstract}
We perform several calculations that, while not essential to the main paper, support the findings contained therein.
We first of all perform a `stress test' on several anomalous dispersion relations to investigate rigorously the limitations of the Gaster series with respect to its convergence.  We then discuss the complex-analytic properties of the dispersion relation that limit its convergence.  We find that branch cuts on the real line are a severe limitation, while on the other hand other anomalous situations (such as multiple dominant saddle points) can be handled by the series representation.
\end{abstract}

\setcounter{tocdepth}{1}
\tableofcontents

\newpage

\section{Inviscid shear flow (Juniper)}
\label{sec:juniper}

The dispersion relation for inviscid shear flow in a confined geometry is studied (see~\citet{Juniper2006}).  For the varicose mode, in a non-dimensional framework, this is given in implicit form (\citet{Juniper2006}, p. 174) as
\begin{equation}
S\left(1+\Lambda-\frac{\omega}{k}\right)^2\coth(k)+\left(1-\Lambda-\frac{\omega}{k}\right)^2-k\Sigma=0,
\label{eq:disp_implicit}
\end{equation}
where $(S,\Lambda,h,\Sigma)$ are parameters, $\omega$ is the frequency, and $k$ is the wavenumber.  This can be solved to give $\omega$ explicitly as a function of $k$:
\begin{equation}
\omega=\frac{k\left[2\coth(k)\pm \sqrt{[\coth k+\coth (kh)]k-4\coth (k)\coth (kh)}\right]}{\coth k+\coth kh}.
\label{eq:disp_explicit}
\end{equation}
Evidently, this dispersion relation contains simple poles at those values of $k$ that solve $\coth k+\coth kh=0$; this can be solved to give
\begin{equation}
k=\imag\frac{n\pi}{1+h},\qquad n=1,2,\cdots
\label{eq:poles}
\end{equation}
(there is a superficial singularity at $n=0$).
Furthermore, the function is non-differentiable when at the point where the radicand is zero: this is the real value $x$ for which
\[
\left[\coth x+\coth (xh)\right]x-4\coth (x)\coth (xh)=0,
\]
with $x=x_0\approx 2.05$ in the right half-plane $\Re(k)>0$ (RHP).  
Juniper takes a branch cut in the RHP along the line $[x_0,\infty)$, such that the square root in Equation~\eqref{eq:disp_explicit} is understood in the usual way as having a branch cut that extends from the point of non-differentiability to $+\infty$. 
%
%
%
%
%
However, further points along the imaginary axis exist where the radicand vanishes: these produce branch cuts that co-exist with the poles~\eqref{eq:poles} along the imaginary axis.
%
%

The dispersion relation (e.g. Equation~\eqref{eq:disp_explicit}) possesses the symmetry
\[
\omega(-k^*)=-\omega^*(k),
\]
meaning that $\omi(LHP)=\omi(RHP)$, and $\omr(LHP)=-\omi(RHP)$, where `LHP' means `left half-plane'.  However,~\citet{Juniper2006} cuts the dispersion relation in two along the imaginary axis, and takes the positive sign in Equation~\eqref{eq:disp_explicit} in $\Re(k)>0$ and the negative sign in the LHP $\Re(k)<0$.  This results in a dispersion relation that varies smoothly as the imaginary axis is traversed.  The result of this procedure is Figure~2 in the paper of Juniper.  

\begin{figure}[htb]
\centering
\includegraphics[width=0.8\textwidth]{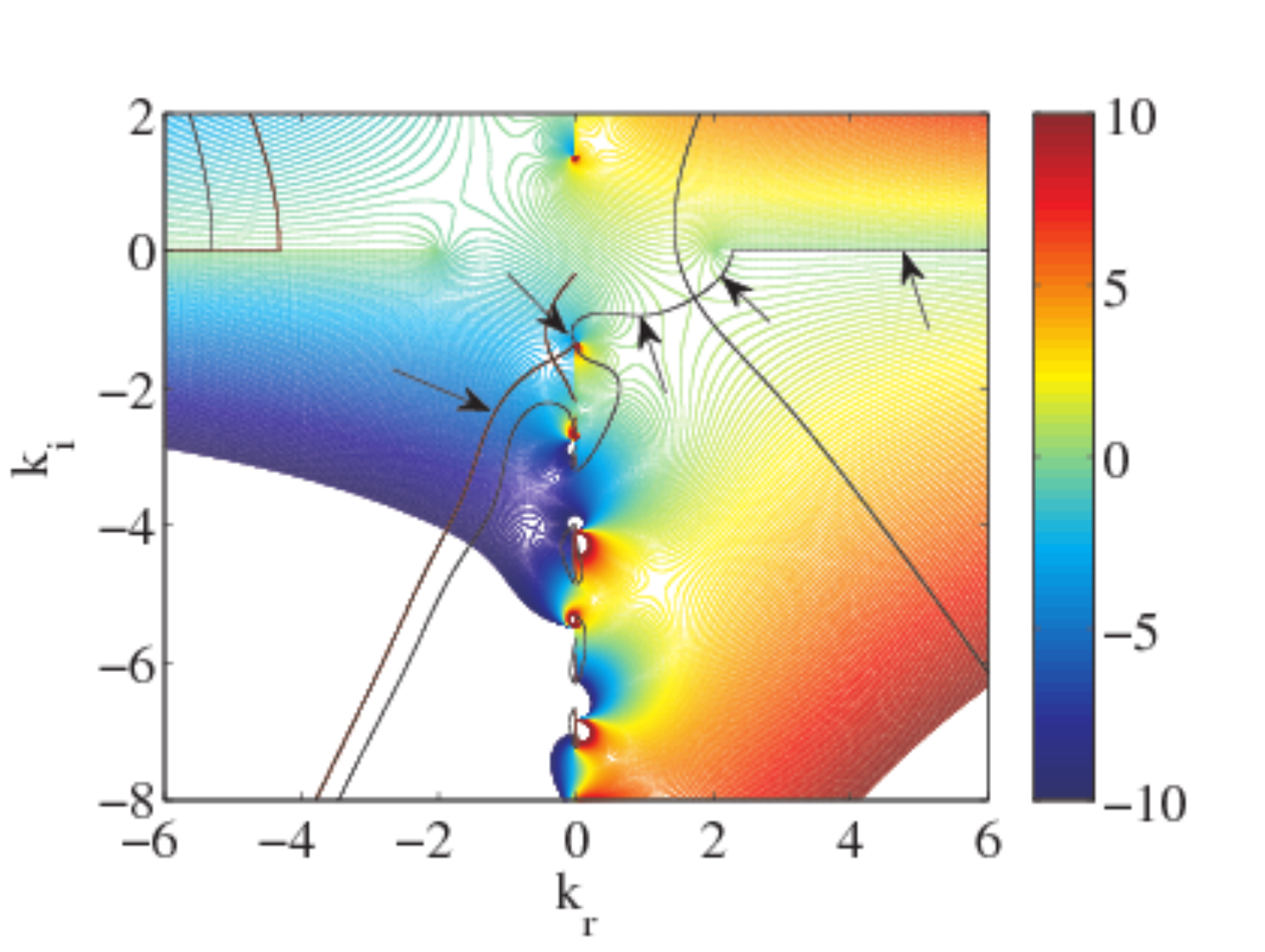}
\caption{Reproduction of Figure~2 in the paper of~\citet{Juniper2006}.  The steepest-descent part of the contour $\omr= 1.50564$ is shown with the arrows. }
\label{fig:juniper1}
\end{figure}

\begin{figure}[htb]
\centering
\includegraphics[width=0.5\textwidth]{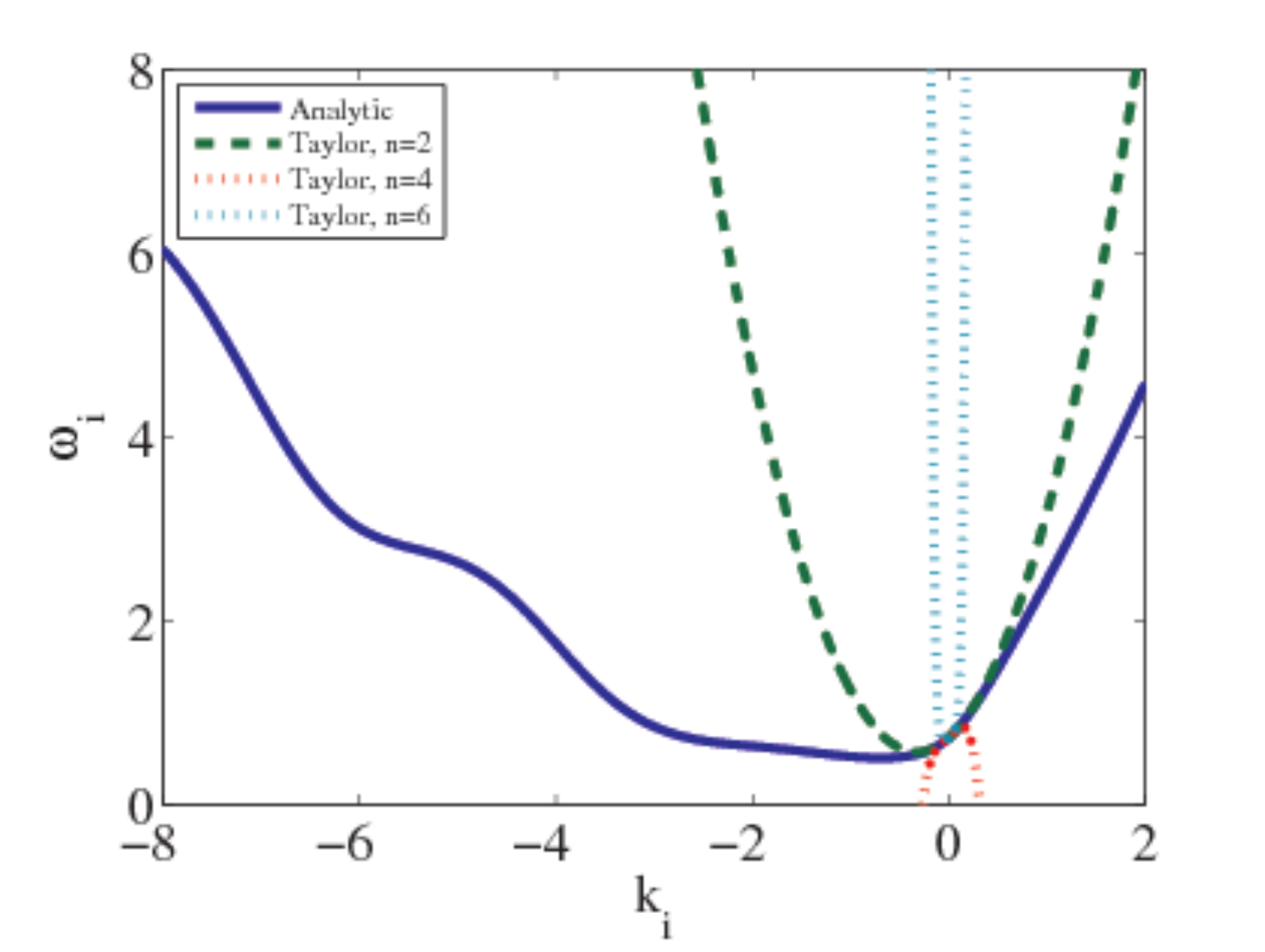}
\caption{Taylor polynomials based on the extension of Gaster's formula at the saddle-point location.}
\label{fig:juniper2}
\end{figure}

The aim of this section is to study the extent to which the power-series dispersion relation based along the real axis (referred to herein as the Gaster series) reproduces the dynamically-relevant saddle points of the dispersion relation.  The analytic dispersion relation is generated in Matlab according to the procedure outlined above.  Dynamically relevant saddle points are identified via the Briggs criterion.  In the RHP, for the parameters $(S,\Lambda,h,\Sigma)=(1,1,1.3,1)$, only one such saddle exists, at $k\approx (1.72,-0.682)$.  The steepest-descent path is therefore that contour of $\omr$ that passes through the saddle, along that path along which $\omi$ attains a minimum.  This corresponds to $\omr\approx 1.50564$.  The results of this calculation are shown in Figure~\ref{fig:juniper1}.
It is of interest to attempt to apply the analytic extension of Gaster's formula to the dynamically-relevant saddle point at $k\approx (1.72,-0.682)$.  This amounts to a Taylor series centred at $(1.72,0)$.  However, such a series will be hampered by the presence of the branch point at $(2.05,0)$, and its radius of convergence is thus $R\apprle 0.33$.  The disc of convergence $D((1.72,0),R=0.33)$ does not include the dynamically-relevant saddle point (indeed, it fails in a substantial way to include this point), and the series will therefore not converge at the saddle point.  We examine further  whether a Taylor polynomial based on the extension of Gaster's formula (but which does not converge to the generating function) yields a good approximation to the saddle point (Figure~\ref{fig:juniper2}).
Clearly, the second truncation is best, while all other truncations fail badly (the odd-numbered polynomials do not yield any improvements with respect to the even polynomials and are not included here).  However, even the second truncation fails to reproduce the saddle point in any meaningful way.  This is not surprising, given the non-convergence of the Gaster series at the saddle point.  However, the result is still important, since it highlights a limitation of our formula.  

Finally, we highlight a key difference between this model problem and that considered in the main paper: in the main paper, the Gaster series of the dispersion relation produced Taylor polynomials that gave an excellent approximation to the saddle point, in spite of the fact that the saddle was located beyond the disc of convergence.  This was because the saddle was just barely outside the disc of convergence.  On the other hand, in this example, the saddle is located at a distance greater than $2R$ away from the series centre, where $R$ is the series' radius of convergence.  The presence of a branch point on the real axis hinders an application of our formula, or even the truncated version thereof.

\section{Eady model (Brevdo)}
The dispersion relation for a simplified model of baroclinic instability (Eady model) is studied (see~\citet{Brevdo1988}).  The exact dispersion relation corresponding to the discrete spectrum is known (e.g. Equation~(4.18) in the paper of~\citet{Brevdo1988}):
\begin{equation}
\omega(k)=\tfrac{1}{2}k\pm \sqrt{\frac{k^2}{\mu^2}\left(\tfrac{1}{2}\mu-\coth \tfrac{1}{2}\mu\right)\left(\tfrac{1}{2}\mu-\tanh\tfrac{1}{2}\mu\right)},\qquad \mu^2=(k^2+\ell_m^2)S
\label{eq:eady1}
\end{equation}
where $(S,\ell_m)$ are parameters taken to be $(0.25,\pi/2)$.  
We prove the following result:

{\proposition{The function 
\begin{equation}
\Phi(k):=\left(\tfrac{1}{2}\mu-\coth\tfrac{1}{2}\mu\right)\left(\tfrac{1}{2}\mu-\tanh\tfrac{1}{2}\mu\right),\qquad \mu=\sqrt{(k^2+\ell_m^2)S}
\label{eq:phidef}
\end{equation}
contains only isolated singularities.  Moreover, $\mu=0$ is not a singular point.}}

\vspace{0.1in}
\noindent {\textbf{Proof}} We re-write $\coth(z)=z^{-1}+zf_C(z)$, where $f_C(z)$ is a function of $z^2$ with a Taylor series around $z=0$; in particular,
\begin{subequations}
\begin{equation}
f_C(z)=\tfrac{1}{3}-\tfrac{1}{45}z^2+\tfrac{2}{945}z^4+\cdots.
\end{equation}
Similarly, write $\tanh (z)=z+z f_T(z)$, where $f_T(z)$ is also a function of $z^2$ with a Taylor series around $z=0$; in particular,
\begin{equation}
f_T(z)=-\tfrac{1}{3}z^2+\tfrac{2}{15}z^4+\cdots.
\end{equation}%
\label{eq:taylorft}%
\end{subequations}%
Then, taking $z=\mu/2$, the function $\Phi(k)$ can be written as $\Phi(k)=f_T(z)(1-z^2)+z^2F_T(z)F_C(z)$, where $\mu^2=(k^2+\ell_m^2)S$.  Thus, $k$ appears in $\Phi(k)$ only in the form $\mu^2$.  Moreover, the form of Taylor series~\eqref{eq:taylorft} demonstrates that $\Phi(k)\sim -\mu^2/3$ as $\mu\rightarrow 0$.  \qed
\vspace{0.1in}

\begin{figure}[htb]
\centering
\includegraphics[width=0.8\textwidth]{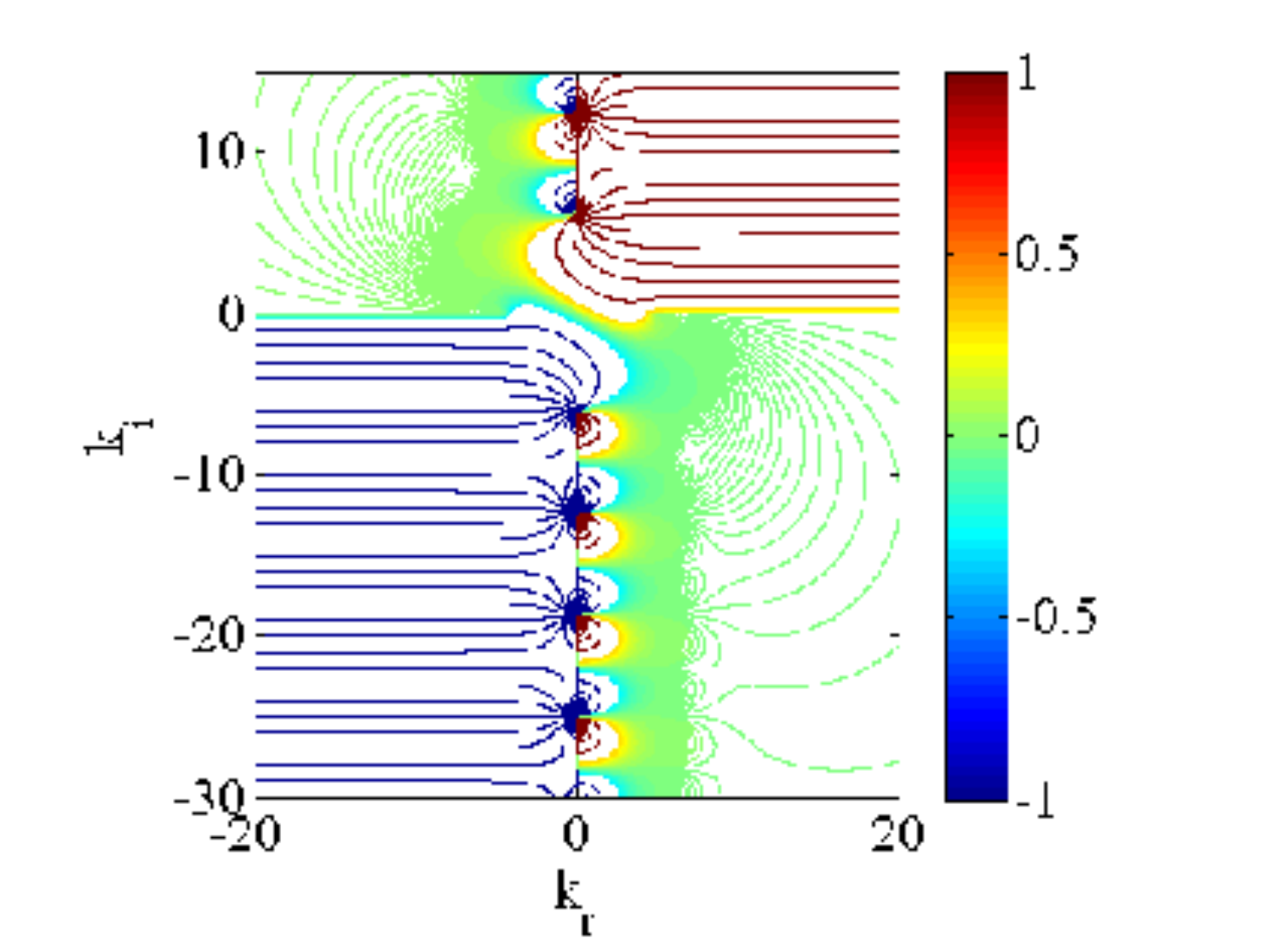}
\caption{Complex dispersion relation for the Eady model, $\omi(k_\mathrm{r},k_\mathrm{i})$ }
\label{fig:brevdo1}
\end{figure}

\noindent We now list the analytic properties of Equation~\eqref{eq:eady1}:
\begin{enumerate}
\item Since $\Phi(\mu)\sim -\mu^2/3$ as $\mu\rightarrow 0$, the dispersion relation $\omega(k)=(k/2)\pm \sqrt{k^2\Phi(\mu)/\mu^2}$ admits no pole at $\mu^2=0$.
\item The only possible poles in the dispersion relation therefore occur when $\coth(\mu/2)=\infty$, or when $\tanh(\mu/2)=\infty$.  This happens when $\mu/2\imag$ is half-integral (the $\tanh$-function), or full-integral (the $\coth$-function), hence 
\[
\mu/2=\imag (\pi/2,3\pi/2,5\pi,\cdots),\text{ or }\mu/2=\imag (\pi,2\pi,3\pi,\cdots).
\]
In other words, the singularities are located at
\begin{equation}
k=\imag \sqrt{\frac{4n^2\pi^2}{S}+\ell_m^2},\qquad k=\imag \sqrt{\frac{4\pi^2}{S}\left(n+\tfrac{1}{2}\right)^2+\ell_m^2}.
\label{eq:eady_sing}
\end{equation}
(The value $n=0$ is not a singularity, since point (1) confirms that $\mu=0$ is a regular point.)

\item There are branch points along the real axis at $k=x_0\approx \pm 4.543$,  corresponding to real $k$-values where $\Phi(\mu)=0$.  Further branch points exist along the imaginary axis corresponding also zeros of $\Phi(\mu)$.
\end{enumerate}
Using this information, we construct a dispersion-relation manifold  as in the paper of~\citet{Juniper2006} by placing branch cuts along the real axis at $[x_0,\infty)$ and $(-\infty,-x_0]$.  Further branch cuts along the imaginary axis are placed in such a way as to connect the singularities in Equation~\eqref{eq:eady_sing}.  Finally, again as in the paper of~\citet{Juniper2006}, the manifold is rendered smooth across the axis $\Re(k)=0$ by joining together two sheets corresponding to the $\pm$ sign in Equation~\eqref{eq:eady1} (obviously, the smoothness does not extend to the singularities or branch cuts along the imaginary axis).  
This procedure amounts to picking the following solution branch of Equation~\eqref{eq:eady1}:
\begin{equation}
\omega(k)=\tfrac{1}{2}k+ \,\mathrm{sign}[\Re(k)]k \sqrt{\frac{1}{\mu^2}\left(\tfrac{1}{2}\mu-\coth \tfrac{1}{2}\mu\right)\left(\tfrac{1}{2}\mu-\tanh\tfrac{1}{2}\mu\right)},
\label{eq:eady1}
\end{equation}
where the $\sqrt{\cdot}$ function is understood in the usual sense, as having a branch cut that extends from the point of non-differentiability to $+\infty$.
The results of this analysis are plotted in the dispersion relation in Figure~\ref{fig:brevdo1}.
There is a family of saddles that arise from the singularities of the dispersion relation along the imaginary axis.  As demonstrated by~\cite{Brevdo1988}, none of these saddles satisfies the Briggs criterion, and the flow is therefore only convectively unstable.  This is closely analogous to the `confinement saddles' in Section~\ref{sec:juniper}.  A further analogy with Section~\ref{sec:juniper} is the branch point on the real axis which hinders the development of a Gaster series to connect with the saddle point at complex $k$.  Given our experience in Section~\ref{sec:juniper}, where a branch point on the real axis hindered the development of such a series, we do not attempt to compute such a series here.


\section{Falling films (Brevdo et al.)}
\label{sec:dias}

In the work by~\citet{Brevdo1999}, the stability of a viscous liquid film on an inclined plane is studied with respect to perturbations to the film free surface.  The problem contains three independent parameters: the inclination angle $\theta$, the Reynolds number $Re$, and the Weber number $W$.  For a wide class of physically-relevant parameters, the system is found only to be convectively unstable.  To demonstrate these stability properties, the authors studied a modified dispersion relation
\[
\omega_V(k)=\omega(k)-kV,
\]
where $\omega(k)$ is the usual dispersion relation got by solving the Orr--Sommerfeld equation with the relevant free-surface boundary conditions, and $V$ is a free parameter.  Physically, the parameter $V$ is a velocity measured with respect to the laboratory frame: the quantity $\omega_V(k)$ can therefore be thought of as the wave frequency observed in a moving frame of reference.    In the presence of a single dominant saddle, the asymptotic growth rate of the pulse is
\[
\sigma(V)=\Im\left[\omega_V(k_*)\right],\qquad \frac{d\omega_V}{dk}\bigg|_{k_*}=0,
\]
provided the saddle point pinches according to the Briggs criterion.
\citet{Brevdo1999} studied the function $\sigma(V)$ for a wide class of parameters $(\theta,Re,W)$ and found $\sigma(0)<0$ always, indicating convective instability.  Moreover, they found that the function $\sigma(V)$ has two branches, due to the presence of two equally-dominant saddle points in the complex function $\omega_V(k)$. 
A previous approach to computing $\sigma(V)$ was based on the assumption that the modified dispersion relation $\omega_V(k)$ should contain only one dominant saddle~\citep{Monkey1988,Deissler1987}; the findings in the paper of~\citet{Brevdo1999} therefore demonstrate that this approach is limited to situations where only one saddle point is of interest.

\begin{figure}[htb]
\centering
\includegraphics[width=0.8\textwidth]{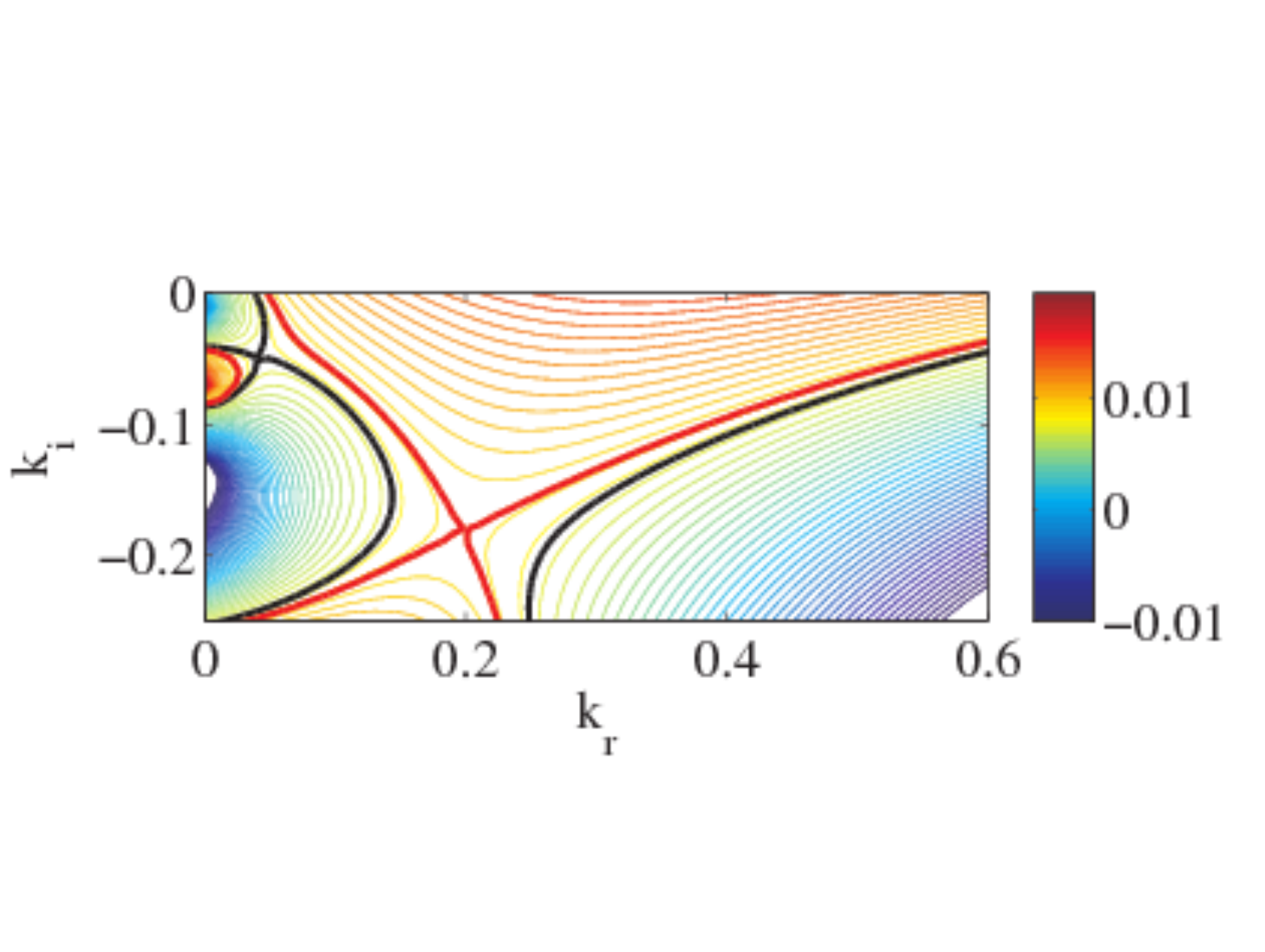}
\caption{The modified dispersion relation $\omega_V(k)$ at $V=1.16$, corresponding to Figure~5 in the paper of~\citet{Brevdo1999}.  The bold lines show the collision of the different spatial branches that occur at the two saddles.}
\label{fig:disp_dias}
\end{figure}

\begin{figure}[htb]
\centering
{\includegraphics[width=0.45\textwidth]{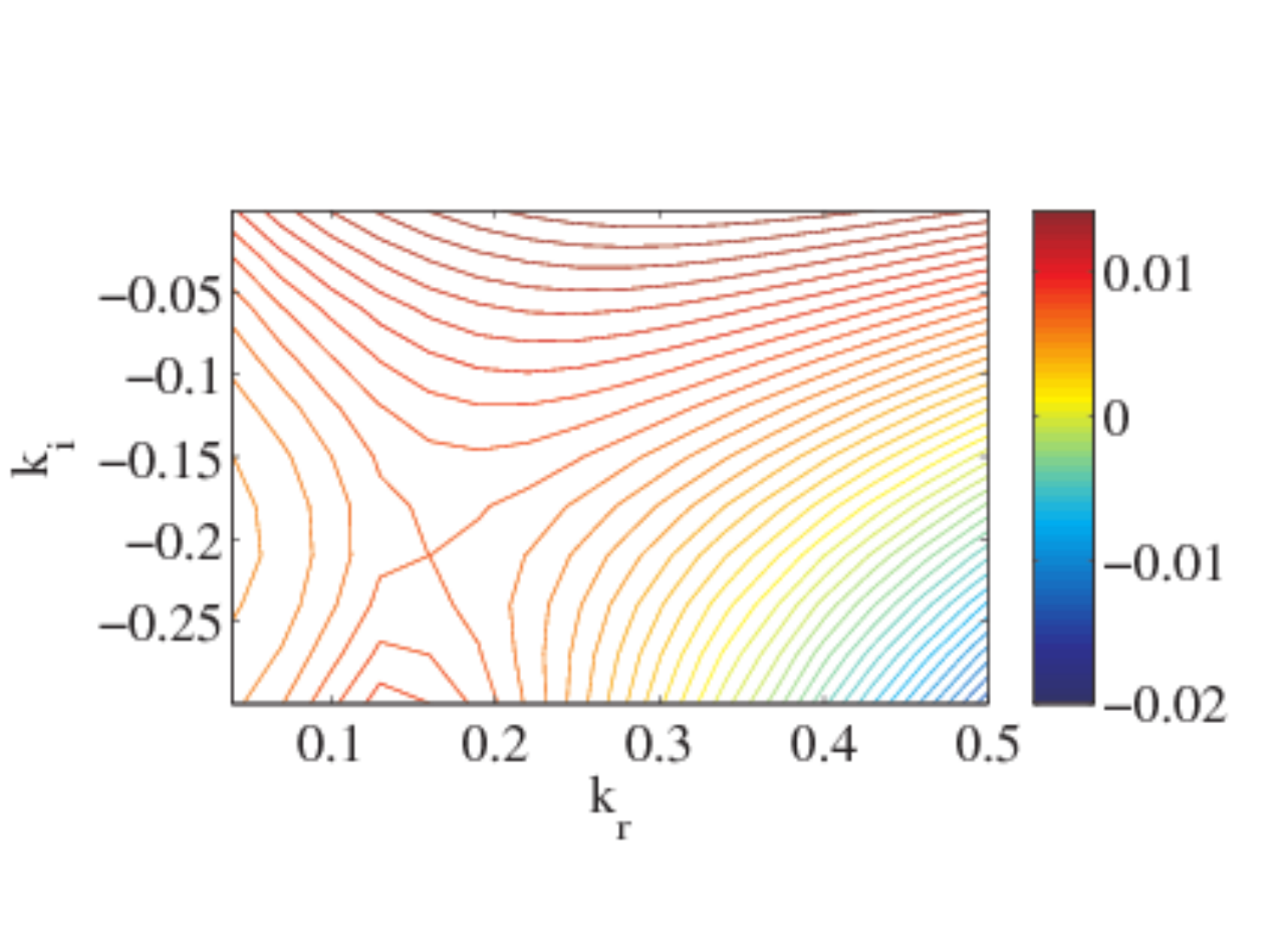}}
{\includegraphics[width=0.45\textwidth]{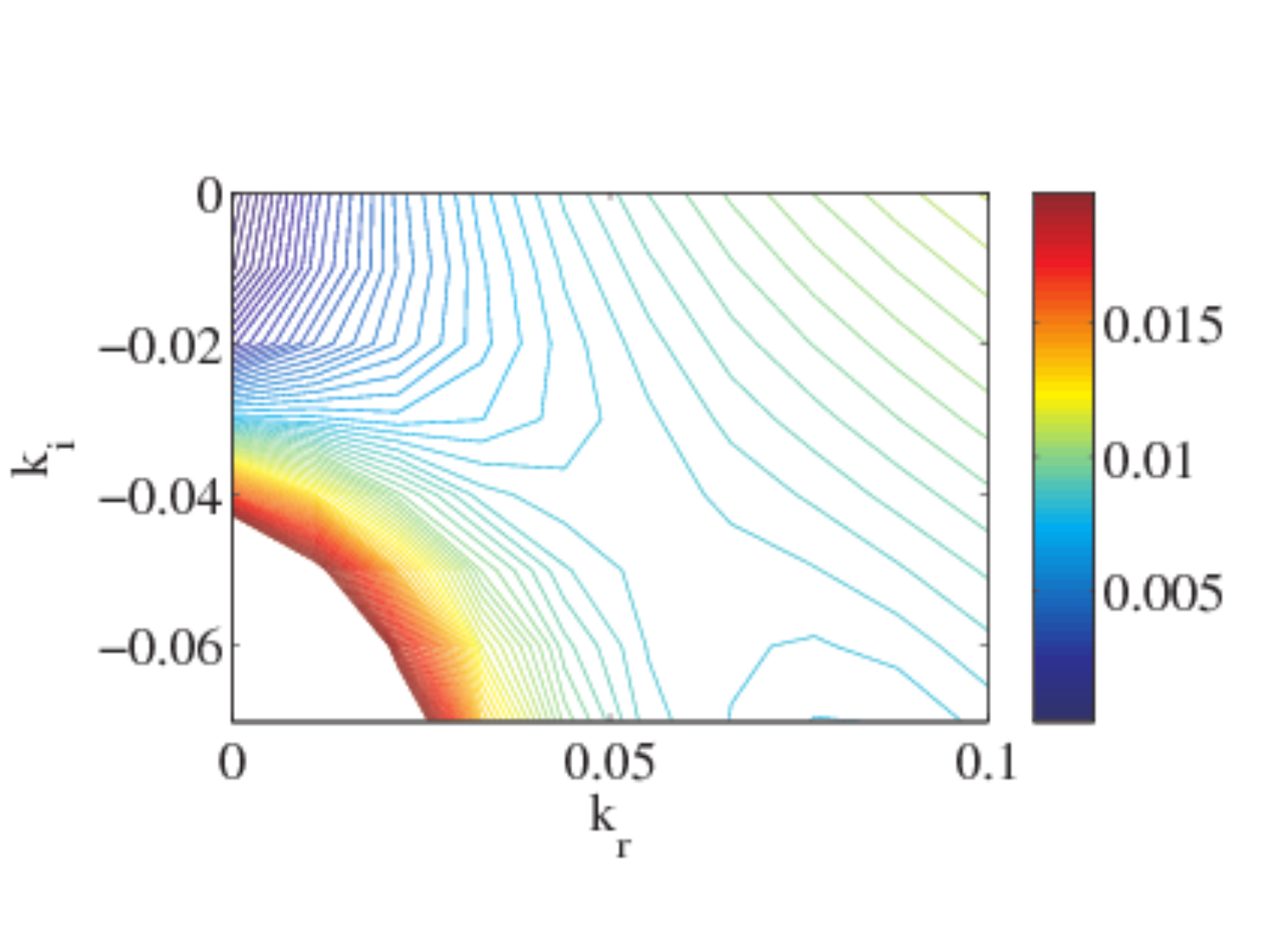}}
\caption{Cubic approximation to the saddle points.}
\label{fig:approx_landscape}
\end{figure}

It is of interest to know whether a Gaster series  can reproduce the anomalous behaviour described in the work of~\citet{Brevdo1999}.  For that reason, we implemented an Orr--Sommerfeld analysis of the falling-film equations for parameter values $(\theta,Re,W)=(4.6^{\mathrm{o}},200,14.18)$ (as in Figure~5 in the work of~\citet{Brevdo1999}), and plotted the resulting complex dispersion relation (Figure~\ref{fig:disp_dias} herein).
The saddles are estimated as in Table~\ref{tab:disp_dias} (a more refined estimate, in agreement with the work of~\cite{Brevdo1999} can be obtained by increasing the resolution of the numerical scan through complex $k$-space in the Orr--Sommerfeld solver).  Because the real dispersion relation $\omega(k_\mathrm{r})$ has a cusp at $k_\mathrm{r}=0$, the convergence of the Gaster series is restricted to a disc centred on the point of interest $(k_\mathrm{r},0)$, of radius $k_\mathrm{r}$.  Thus, the two saddle points in Table~\ref{tab:disp_dias} lie just barely outside the relevant discs of convergence.  However, because we are just barely outside the region of convergence, as in the main paper, a truncated series will give a good approximation of the true saddle.  
\begin{table}[htb]
\centering
\begin{tabular}{|p{0.6cm}|p{2.5cm}|p{2.5cm}|p{3cm}|p{3.5cm}|}
\hline
    & Numerics, Main Saddle     & Series, Main Saddle     & Numerics, Secondary Saddle       &  Series, Secondary Saddle  \\     
\hline 
\hline
$\omi $            & 0.0086    & 0.0080   & 0.0078    &  0.0080   \\
$k_{\mathrm{r}} $  &  0.20     & 0.16    &  0.045    &   0.055   \\
$k_{\mathrm{i}}$   & -0.18    & -0.21    & -0.048    &  -0.040   \\
\hline
\end{tabular}
\caption{Saddle point at $V=1.16$ computed  directly via a solution of the Orr--Sommerfeld equation, and by a Gaster series.}
\label{tab:disp_dias}
\end{table}
This is  demonstrated in Tab.~\ref{tab:disp_dias} and in Figure~\ref{fig:approx_landscape}.  In these cases, the Gaster series is truncated at cubic order, and the coefficients of the series expansion are computed from the real dispersion relation using centred differences.
Thus, the Gaster series captures -- albeit in a slightly crude way -- the anomalous existence of two `branches' of the growth-rate curve $\sigma(V)$ described in the work of~\citet{Brevdo1999}.  

We anticipate that the entire curve $\sigma(V)$ can be reconstructed -- to a reasonable level of approximation, using truncations of the Gaster series that are `just barely non-converging'.  However, we do not pursue this matter here.  Instead, we focus finally on an aspect of the cubic truncation of the Gaster dispersion relation that highlights its usefulness as an explanatory tool.  In this truncation,  the Gaster series described in the main paper  reduces to
\begin{equation}
\omi(\kr,\ki)=\omitemp(\kr)+\cg(\kr)\ki-\tfrac{1}{2}\frac{d^2\omitemp}{d\kr^2}\ki^2-\tfrac{1}{6}\frac{d^2\cg}{d\kr^2}\ki^3.
\label{eq:omi_approx}
\end{equation}
The necessary saddle-point conditions for instability in a moving frame of reference are $\partial\omi/\partial \ki=V$ and $\partial\omi/\partial \kr=0$; under the approximation~\eqref{eq:omi_approx} these conditions become
\begin{equation}
\frac{d\omitemp}{d\kr}+\frac{d\cg}{d\kr}\ki-\tfrac{1}{2}\frac{d^3\omitemp}{d\kr^3}\ki^2-\tfrac{1}{6}\frac{d^3\omitemp}{d\kr^3}\ki^3=0,
\qquad
\cg(\kr)-\frac{d^2\omitemp}{d\kr^2}\ki-\tfrac{1}{2}\frac{d^2\cg}{d\kr^2}\ki^2=V.
\label{eq:ai_linear}
\end{equation}
%
%
Taking the second condition, we get
\begin{equation}
\ki=\left[-\frac{d^2\omitemp}{d\kr^2}\pm\sqrt{\left(\frac{d^2\omitemp}{dkr^2}\right)^2-2\frac{d^2\cg}{d\kr^2}(V-\cg)}\right]\bigg\slash \frac{d^2\cg}{d\kr^2}:=K_{\pm}(V,\kr),
\label{eq:ai_linear1}
\end{equation}
Substitution into the first condition yields
\begin{equation}
\frac{d\omitemp}{d\kr}+\frac{d\cg}{d\kr}K_{\pm}(V,\kr)-\tfrac{1}{2}\frac{d^3\omitemp}{d\kr^3}\left[K_{\pm}(V,\kr)\right]^2-\tfrac{1}{6}\frac{d^3\omitemp}{d\kr^3}\left[K_{\pm}(V,\kr)\right]^2=0.
\label{eq:saddle}
\end{equation}
For small $V$-values, Equation~\eqref{eq:ai_linear1} has two real roots (`branches' in the terminology of~\citet{Brevdo1999}).  This is precisely the region of parameter space where~\citet{Brevdo1999} observed anomalous behaviour in the $\sigma(V)$-function, with the appearance of two branches of the spatio-temporal growth rate.  These two branches are substituted into Equation~\eqref{eq:saddle} and the precise location of the saddle point is obtained, corresponding to the $\kr$-root of Equation~\eqref{eq:saddle}.  Of course, the computed saddle points should be checked to confirm if they contribute to the impulse response (Briggs criterion).  We would therefore advocate (as in the main paper) that full numerical eigenvalue computations be performed on the one hand, and that the series expansion be considered on the other hand, in order to exploit the complementarity of these two approaches.  The series expansion would be used to explain the precise origin of the multiple dominant saddle points.  Therefore, in conclusion, this twin-track approach demonstrates that the Gaster series approach is applicable to the anomalous $\sigma(V)$-function, and can also be used to explain the source of this anomaly in a straightforward manner.

\section{Spatio-temporal growth rates from direct numerical simulation}

In this section we consider the impulse-response problem
\begin{multline}
\left[U_0(z)\partial_x+\partial_t\right]\left(\partial_z^2+\partial_x^2\right)\psi(x,z,t)-U_0''(z)\partial_x\psi(x,z,t)=Re^{-1}\left(\partial_z^2+\partial_x^2\right)^2\psi(x,z,t),\\
\psi(x,z,t=0)=\delta(z)\delta(x),
\label{eq:impulse}
\end{multline}
with the base state described in Section~5 of the main paper, viz. 
\begin{equation}
U_0(z)=1-\Lambda+2\Lambda\big\{1+\sinh^{2N}\left[z\sinh^{-1}(1)\right]\big\}^{-1},\qquad \Lambda<0,
\label{eq:uzero}
\end{equation}
where $\Lambda$ and $N$ are dimensionless parameters, and $-H<z<H$.  Here $H$ is chosen to be sufficiently large such that confinement has no effect on the linear-stability results.  The purpose of this section is to verify that such a value of $H$ exists.  We do this by   solving the impulse-response problem  via direct numerical simulation (using the method developed by~\citet{Onaraigh2012a}) and by subsequent consideration of the norm
\[
n(x,t)=\left[\int_{-H}^H\left|\psi(x,z,t)\right|^2\,\mathd z\right]^{1/2}.
\]
  The growth rate
\[
\sigma(V)=\lim_{\stackrel{t_1,t_2\rightarrow\infty} {t_2\gg t_1}}\left[\frac{\log n(Vt_2,t_2)-\log n(Vt_1,t_1)+\tfrac{1}{2}\left(\log t_2-\log t_1\right)}{t_2-t_1}\right]
\]
is then extracted (for an explanation of this procedure, see the works by~\citet{Chomaz1998a,Chomaz1998b}).  For a system where the instability is determined by a single dominant mode with a single dominant saddle point, $\sigma(V)$ is a paraboloidal curve whose maximum corresponds to the temporally most-dangerouos growth rate.  Also, the sign of $\sigma(0)$ determines the absolute/convective dichotomy, with $\sigma(0)>0$ for absolute instability.

\begin{figure}[htb]
\centering
\includegraphics[width=0.6\textwidth]{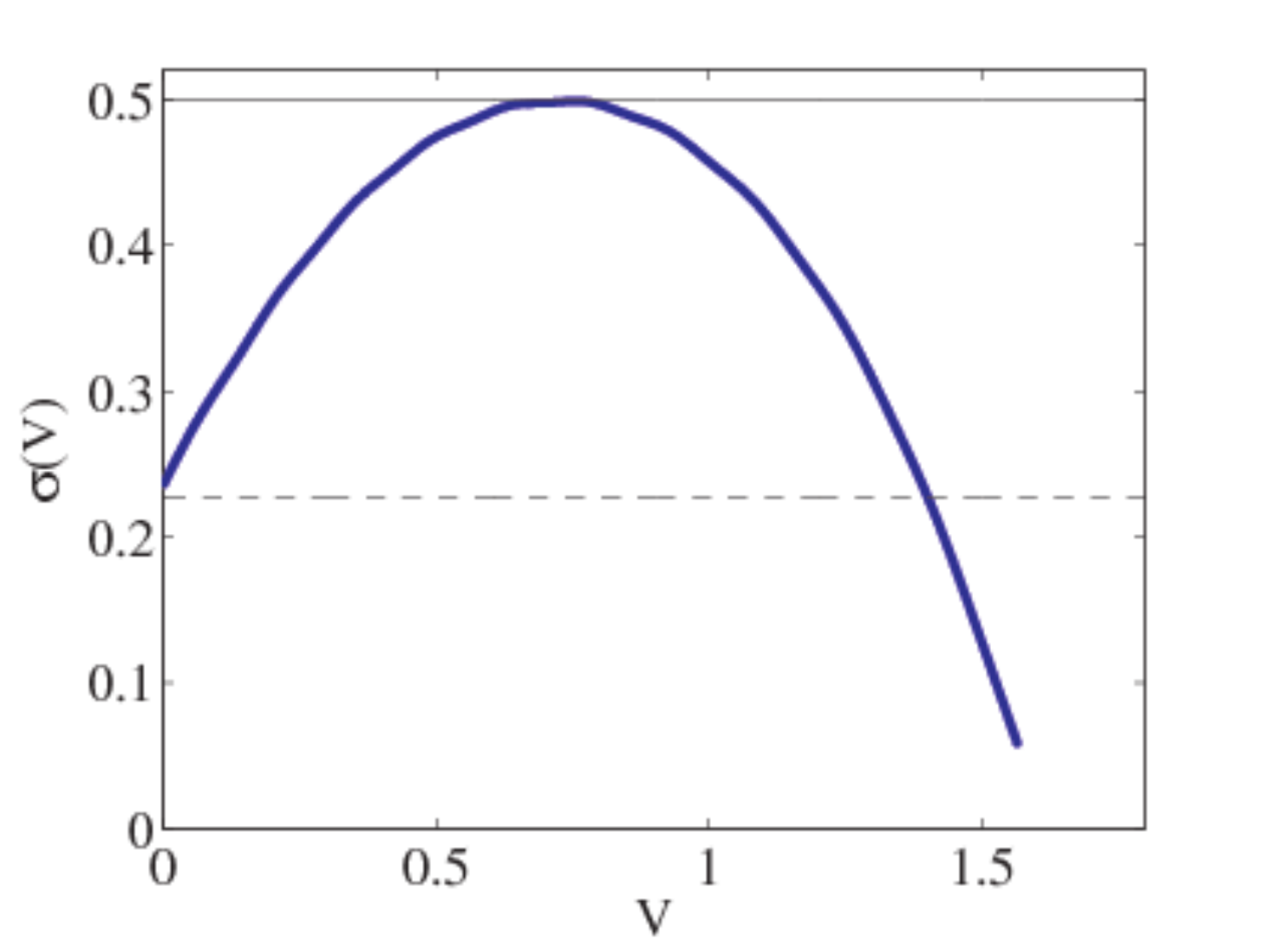}
\caption{Spatio-temporal growth rate computed via direct numerical simulation of Equations~\eqref{eq:impulse}--\eqref{eq:uzero}.  Parameter values: $(Re,\Lambda,N)=(100,-1.1,5)$, and $H=8$.  Solid line: The value of $\omi$ at the temporally most-dangerous mode; Broken line: the value of $\omi$ at the saddle point of $\omega(k)$, in other words, the absolute growth rate.  The values used to draw the lines have been obtained via standard (eigenvalue) stability analysis.  The agreement between the two methods confirms the correctness of the whole approach.}
\label{fig:sigmaV}
\end{figure}

This procedure was applied to Equations~\eqref{eq:impulse}--\eqref{eq:uzero} for parameter values 
\[
(Re,\Lambda,N)=(100,-1.1,5), \text{ and }H=8.
\]
The results are shown in Figure~\ref{fig:sigmaV}.
Clearly, the growth rate possesses only a single branch.  This is in contrast to Section~\ref{sec:dias} of this supplementary report, where $\sigma(V)$ possessed two branches coming from the two contributing saddle points of $\omega_V(k)$.  Additionally, the growth rate of the temporally-most dangerous mode coincides with the maximum of $\sigma(V)$.  Finally, the value of $\omi$ at the saddle point of $\omega(k)$ in the complex $k$-plane coincides with $\sigma(0)$.  These findings confirm that $\sigma(V$) has a single branch and that this branch  corresponds to the most-dangerous mode and its associated saddle in the complex $k$-plane.  In other words,  the confinement saddles have no influence on $\sigma(V)$ for $H$ sufficiently large.

\section{Radius of convergence and optimal truncation}
\label{sec:roc}

In this section we demonstrate that the usual criterion for the convergence of a complex power series applies to the central result of the work, namely the Gaster series
\begin{equation}
\omi(\ar,\ai)=\omitemp(\ar)+\sum_{n=0}^\infty \frac{(-1)^n}{(2n+1)!}\frac{d^{2n}\cg}{d\ar^{2n}}\ai^{2n+1}
+\sum_{n=0}^\infty \frac{(-1)^{n+1}}{(2n+2)!}\frac{d^{2n+2}\omitemp}{d\ar^{2n+2}}\ai^{2n+2}.
\label{eq:omi_taylor}
\end{equation}
To maintain the connection with the main part of the paper, we return the the notation used therein, and denote wavenumbers by the symbol $\alpha$.
It is tempting to conclude that the convergence properties of the result~\eqref{eq:omi_taylor} do not depend on the global topography of the function $\omega(\alpha)$, and that a local analysis (similar to that used in Taylor's theorem in real analysis) applies.  We demonstrate here that this conclusion is incorrect.  We also discuss the notion of an optimal truncation for a divergent power series, such that a finite truncation of a divergent power series can be used to approximate the generating function of the power series.   This result explains the `over prediction' obtained in Section~5 of the main paper.

\begin{figure}[htb]
\centering
\includegraphics[width=0.6\textwidth]{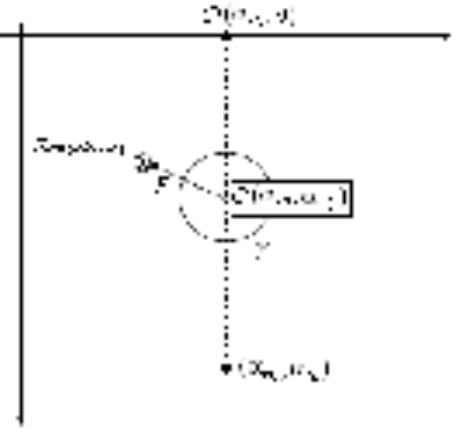}
\caption{Figure showing the centre of the power series, the singularity, and the point of interest at which the series is to be evaluated.}
\label{fig:sketch}
\end{figure}

The starting-point for the convergence analysis is Taylor's theorem applied to the \textit{purely real} Taylor expansion~\eqref{eq:omi_taylor}.  
We introduce some notation (see also Figure~\ref{fig:sketch}):  
%
%
%
%
\begin{enumerate}
\item The point $O=(\alpha_{\mathrm{r}0},0)$: The centre of the power series based on the real axis.
\item  $(\alpha_{\mathrm{r}0},\alpha_{\mathrm{i}0})$: The point of interest at which the power series is to be evaluated.  For our purposes, $\alpha_{\mathrm{i}0}$ will be negative.

\item The point $C=(\alpha_{\mathrm{r}0},\alpha_{\mathrm{i}C})$: A point  used to compute the remainder in Taylor's theorem; here $\alpha_{\mathrm{i}C}\in[0,\alpha_{\mathrm{i}0}]$.

\item  The point $X=\ax$: The singularity closest to $C$.
\end{enumerate}
We shall assume that the derivatives  $\left(\partial^{k}\omi/\partial\ai^k\right)_{\alpha_{\mathrm{r}0}}$ exist for all orders $k\in\{0,1,2,\cdots\}$, on the interval $(0,\alpha_{\mathrm{i}0})$ (the brackets $\left(\cdot,\cdot\right)$ here denote an open interval, and not a point in the complex plane).
Taylor's theorem guarantees that in approximating $\omi(\ar,\ai)$ by a finite truncation based on Equation~\eqref{eq:omi_taylor}, we incur an error
\begin{equation}
R_n=\frac{1}{(n+1)!}\frac{\partial^{n+1}\omi}{\partial\ai^{n+1}}\bigg|_{\alpha_{\imag C}}\alpha_{\mathrm{i}0}^{n+1},\qquad \alpha_{\mathrm{i}C}\in\left[0,\alpha_{\mathrm{i}0}\right].
\label{eq:rem}
\end{equation}
Using Cauchy's theorem, we have
\begin{equation}
\frac{1}{\imag^{n+1}}\frac{\partial^{n+1}\omega}{\partial\ai^{n+1}}\bigg|_{(\alpha_{\mathrm{r}0},\alpha_{\mathrm{i}C})}=\frac{d^{n+1}\omega}{d\alpha^{n+1}}\bigg|_{(\alpha_{\mathrm{r}0},\alpha_{\mathrm{i}C})}=\frac{(n+1)!}{2\pi\imag}\oint_\gamma\frac{\omega(\alpha)}{\left[\alpha-(\alpha_{\mathrm{r}0}+\imag\alpha_{\mathrm{i}C})\right]^{n+2}}\mathd\alpha,
\label{eq:cauchy0}
\end{equation}
where $\gamma$ is a circle centred at $(\alpha_{\mathrm{r}0},\alpha_{\mathrm{i}C})$, of radius $R$.  Here we have switched from a partial derivative to a standard complex-valued derivative, because the function $\omega(\ar,\ai)$ is analytic away from singularities, and the limits calculated in respect of the derivative are independent of approach.  
The radius $R$ is yet to be determined.  
Also,
\begin{equation}
\left|\frac{\partial^{n+1}\omi}{\partial\ai^{n+1}}\bigg|_{(\alpha_{\mathrm{r}0},\alpha_{\mathrm{i}C})}\right|\leq
\left|\frac{\partial^{n+1}\omega}{\partial\ai^{n+1}}\bigg|_{(\alpha_{\mathrm{r}0},\alpha_{\mathrm{i}C})}\right|\leq
\frac{(n+1)!}{R}\left(\max_\gamma |\omega(\alpha)|\right).
\label{eq:cauchy1}
\end{equation}
Combining Equations~\eqref{eq:rem} and~\eqref{eq:cauchy1}, we have
\[
|R_n|\leq \left(\max_\gamma |\omega|\right)\left|\frac{\alpha_{\mathrm{i}0}}{R}\right|^{n+1}.
\]
To make $|R_n|\rightarrow 0$ as $n\rightarrow\infty$, we would like to take $R>|\alpha_{\mathrm{i}0}|$.  However, we cannot take $R$ as arbitrary, we must have $R<|CX|$, where $|CX|$ is the distance between $C$ and the closest singularity thereto.  Because we have no \textit{a priori} knowledge of the location of $C$, we must take account of the worst-case scenario, where $C=O$.  Then, for the series to converge, we must have 
\begin{equation}
R<|OX|,
\label{eq:app:roc}
\end{equation}
 that is, in order for the power series centred at $O$ to converge, the point of interest $(\alpha_{\mathrm{r}0},\alpha_{\mathrm{i}0})$ must be inside a disc of radius $R$, where $R$ is the distance between the power-series centre and the nearest singularity.  In other words, the standard criterion for complex power series applies: the power series is valid provided the point of interest lies within the disc of convergence.

Equation~\eqref{eq:app:roc} is a sufficient condition for the Taylor series for $\omi(\ar,\ai)$ to be valid.  If we extend this Taylor series using the Cauchy--Riemann conditions and construct a Taylor series in $\ai$ for $\omr(\ar,\ai)$, the standard results of complex-variable theory apply, and  Equation~\eqref{eq:app:roc} is both sufficient \textit{and} necessary in order for the Taylor-series pair for $(\omr,\omi)$ to converge.  
On the other hand, it is possible for the Taylor series of $\omi$ to converge beyond the radius of convergence, while at the same time the Taylor series for $\omr$ diverges.  
%
%
%
%
%
However, this rather pathological scenario is unlikely to occur in the applications contained in this paper: for example, Figure~3 in the main paper shows that the finite truncations to the power series for both $\omi$ \textit{and} $\omr$ are good approximations to the underlying dispersion relation outside of the radius of convergence of the power series.  For a more complete description of this pathology, see Section~\ref{sec:roc:path} below.

\begin{figure}[htb]
\centering
\subfigure[]{\includegraphics[width=0.49\textwidth]{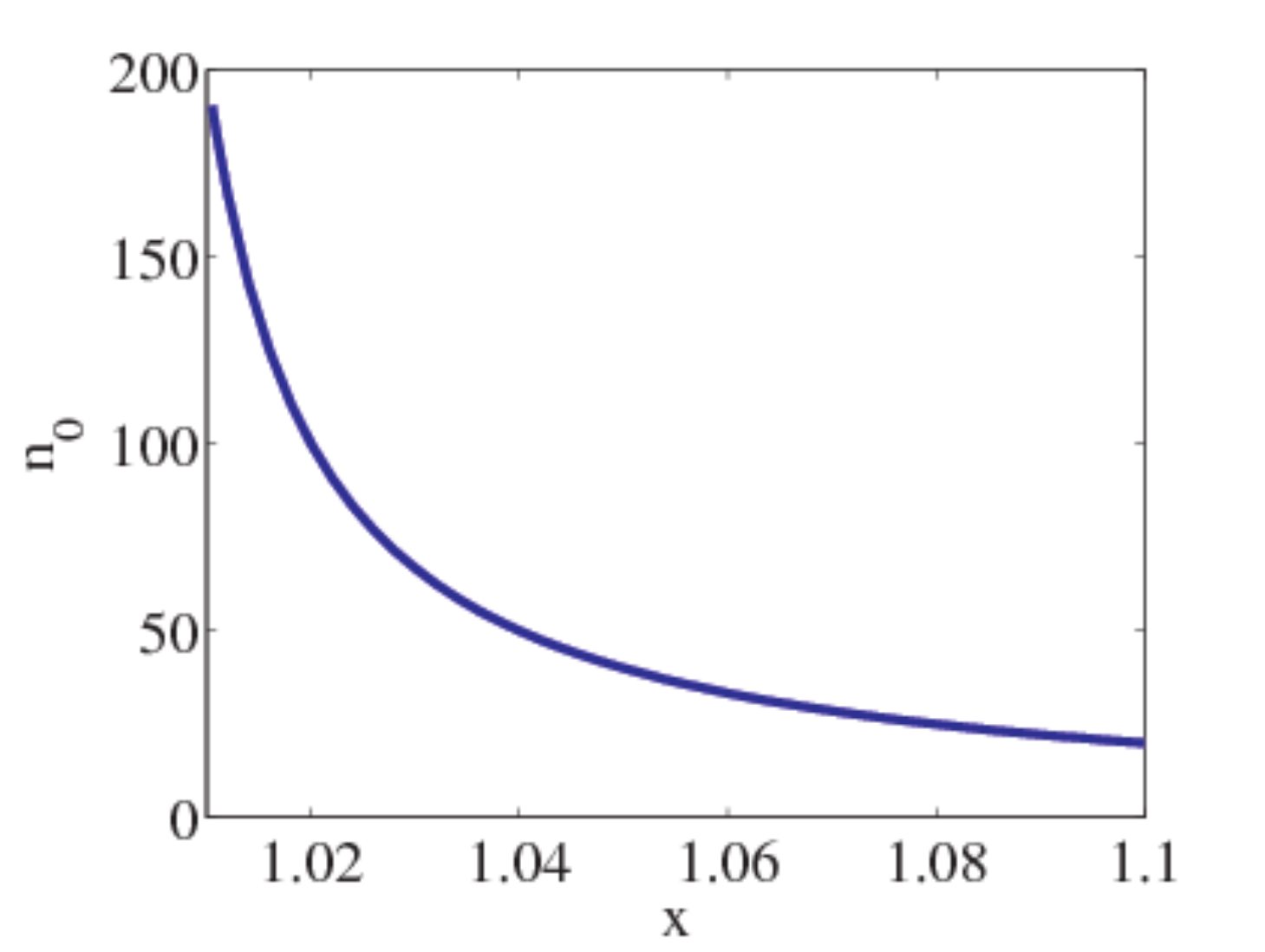}}
\subfigure[]{\includegraphics[width=0.49\textwidth]{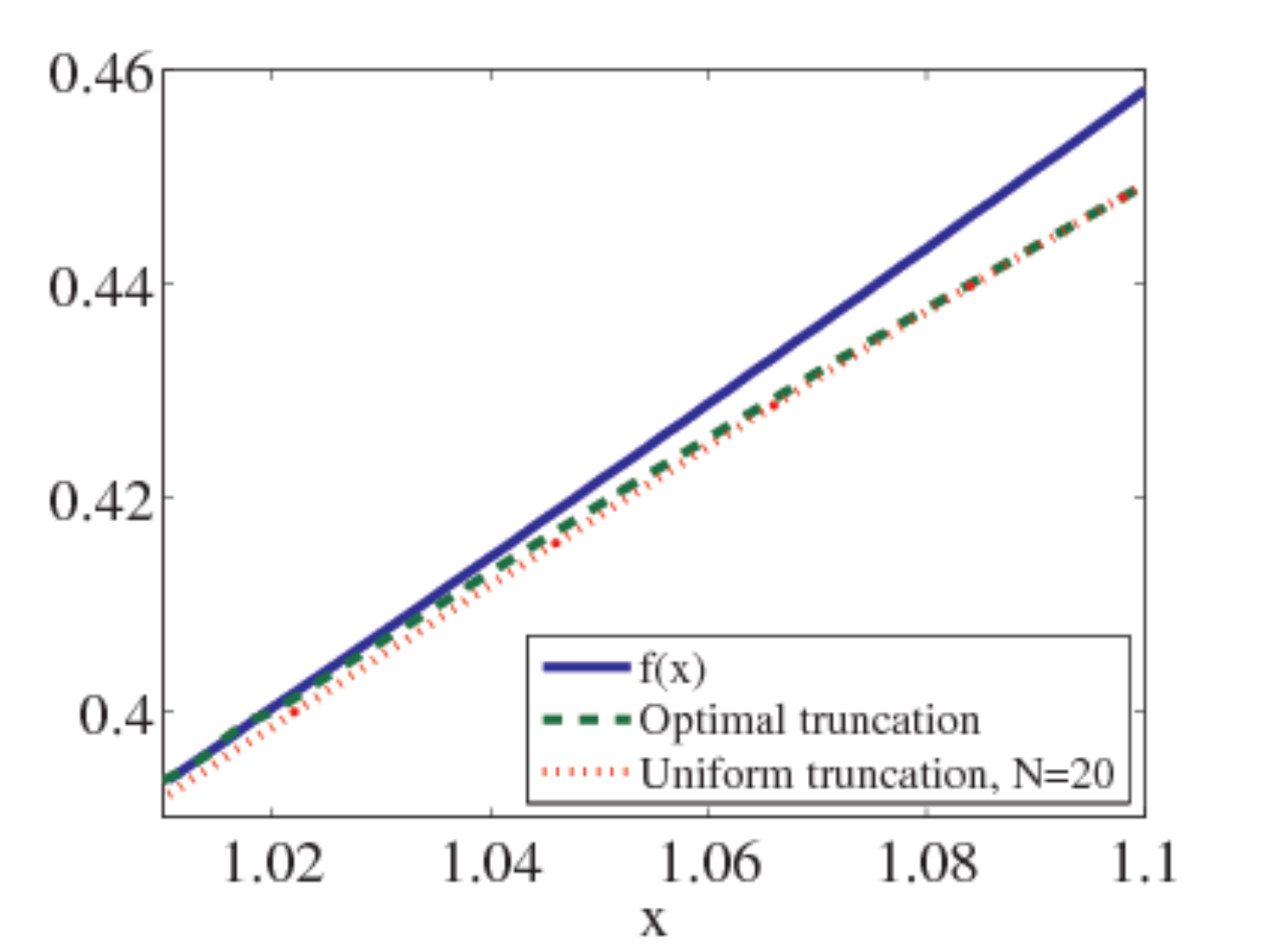}}
\subfigure[]{\includegraphics[width=0.49\textwidth]{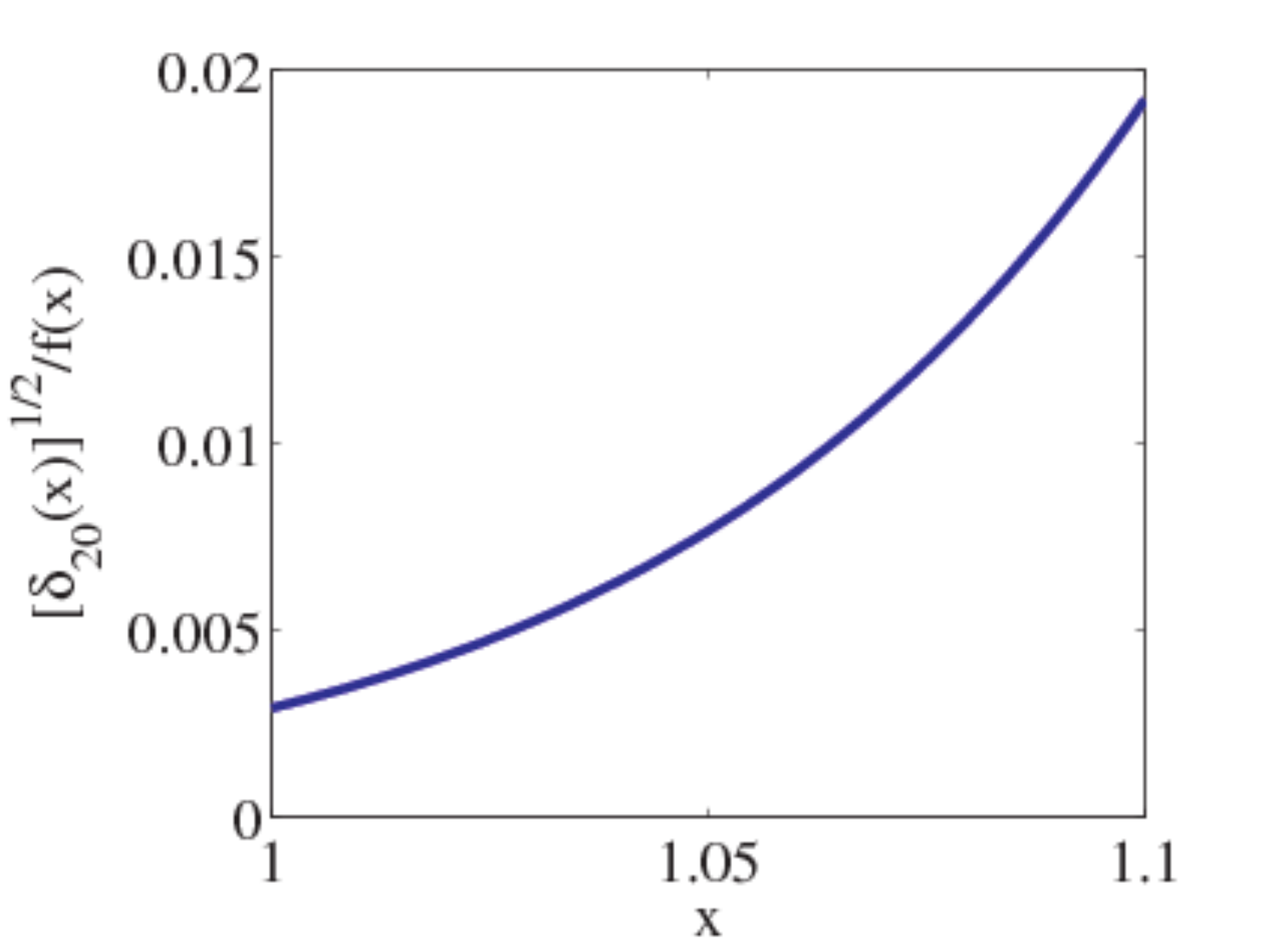}}
\caption{Analysis of finite truncations of the divergent power series~\eqref{eq:app:example_log}.  (a)  The optimal truncation value $\trunc$; (b) The difference between the true function $f(x)=(1+x)\log(1+x)-x$ and the optimal truncation; (c) the relative error in making a uniform truncation.}
\label{fig:app:ot0}
\end{figure}

Our next discussion  concerns the notion of \textit{optimal truncation} for divergent power series.
Consider a real-valued function $f(x)$ that admits a power-series expansion, centred at the origin:
\begin{equation}
f_P(x)=\sum_{n=0}^\infty c_n x^n,\qquad c_n=\frac{f^{(n)}(0)}{n!}.
\label{eq:app:examplef}
\end{equation}
Assume moreover that the power series $f_P(x)$ converges in a finite interval $|x|<R$, and that $f_P(x)=f(x)$ on the same interval.  If $f_P(x)$ remains finite at $x=R$, then it is possible to use a finite truncation of Equation~\eqref{eq:app:examplef} to approximate $f(x)$ in the region $|x|>R$.  In other words, for a given $x$ such that $|x|>R$, there exists a positive integer $\trunc$ such that the difference
\[
\delta_{\trunc}(x):=\left(f(x)-\sum_{n=0}^{\trunc} c_n x^n\right)^2
\]
is minimized.
As an example of this sort, consider the function $f(x)=(1+x)\log (1+x)-x$.  On the interval $|x|<1$, this function can be represented as a power series:
\begin{equation}
f(x)=\sum_{n=1}^\infty (-1)^{n+1} \frac{x^{n+1}}{n(n+1)}.
\label{eq:app:example_log}
\end{equation}
It is straightforward to show that the power series evaluated at $x=1$ converges (the limit comparison test applies).  Thus, we extend finite truncations of the power series beyond $x=1$ to approximate the underlying function $f(x)$.  In doing so, we wish to minimize  the difference $\delta_{\trunc}(x)$.
Using elementary facts about geometric progressions, this difference can be re-written as
\[
\delta_{\trunc}(x):=\left(\int_0^x\mathd x'\int_0^{x'}\mathd x''\frac{1-(-x'')^{\trunc}}{1+x''}-f(x)\right)^2.
\]
The minimum $\trunc$-value satisfies
\begin{equation}
\frac{\partial}{\partial\trunc}\delta_K(x)=0,\qquad \int_0^x\mathd x'\int_0^{x'}\mathd x''\frac{(x'')^{\trunc}\log(x'')}{1+x''}=0.
\label{eq:app:exampleN}
\end{equation}
For $x>1$, Equation~\eqref{eq:app:exampleN} has a solution with positive $\trunc$.  We have plotted the optimal value of $\trunc$ in Figure~\ref{fig:app:ot0}(a).
The difference between the optimal truncation of the divergent power series and $f(x)$ is shown in Figure~\ref{fig:app:ot0}(b).  There is little difference between the optimal truncation and a uniform truncation with $\trunc=20$.  The relative error between the uniform truncation and the true function $f(x)$ is shown in Figure~\ref{fig:app:ot0}(c): this never exceeds $2\%$ on the interval $x\in(1,1.1]$.

\begin{figure}[htb]
\centering
\subfigure[]{\includegraphics[width=0.48\textwidth]{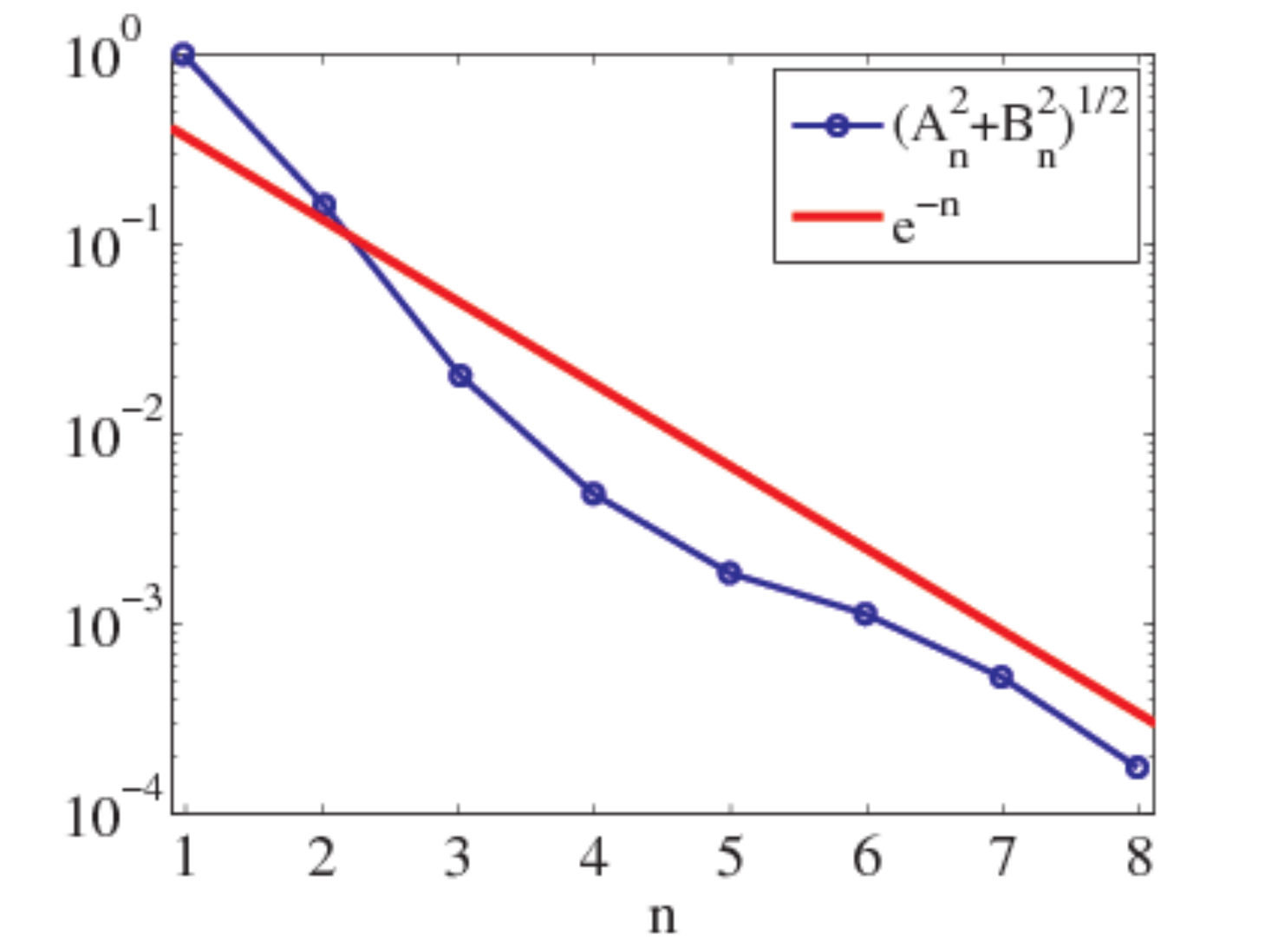}}
\subfigure[]{\includegraphics[width=0.48\textwidth]{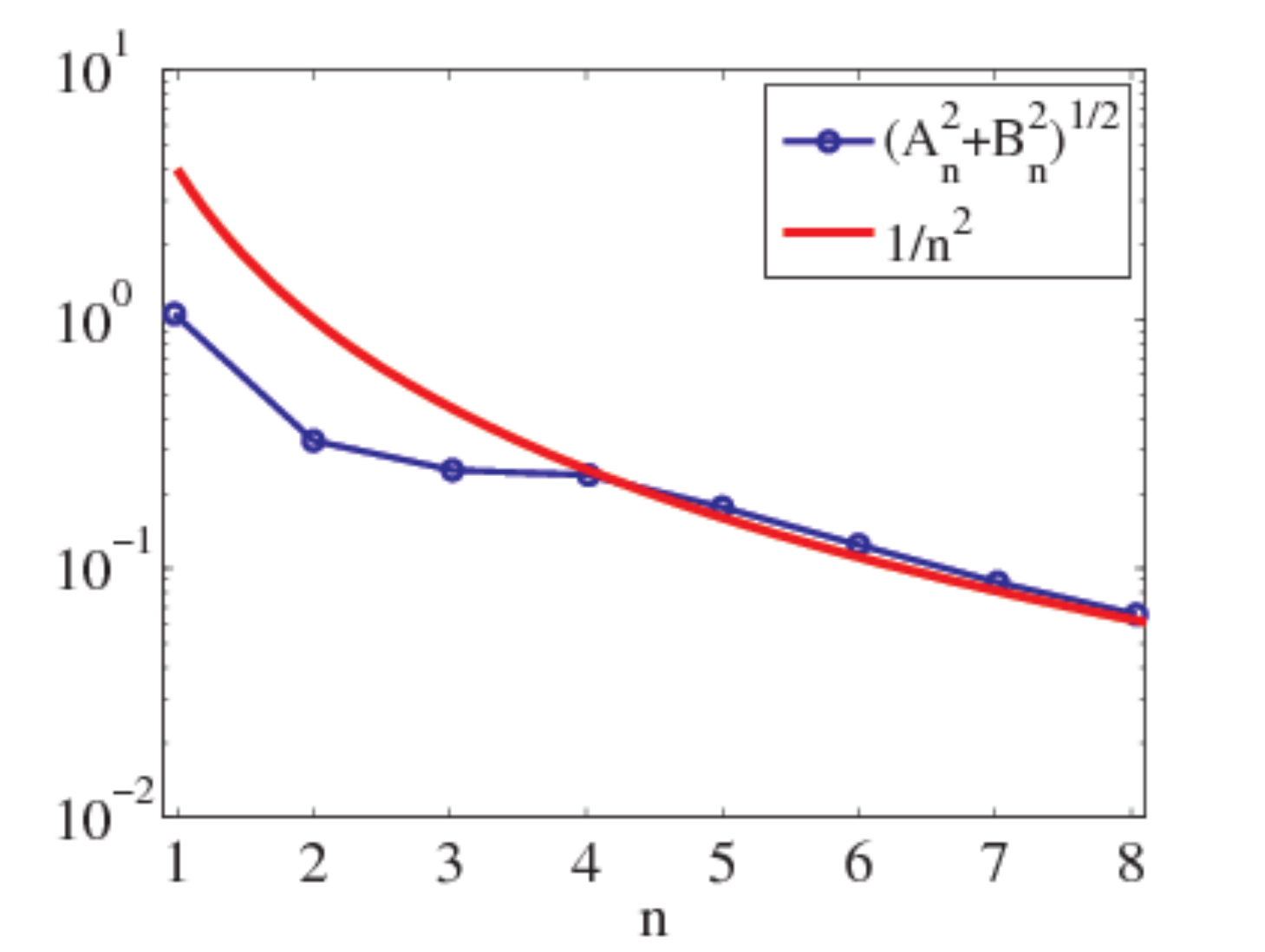}}
\caption{Coefficients of the Taylor series for~\eqref{eq:omi_taylor} ($A_n$) and for the analogous series in $\omr$ ($B_n$), for the model discussed in Section~5 of the main paper.  Here, $Re=100$, and $N=5$.  Case (a) corresponds to $\ar=2.1978$; case (b) corresponds to $\ar=0.7739$ (i.e. a point close to the saddle, for which the radius of convergence is small ($R=0.80$)). }
\label{fig:app:ot}
\end{figure}

Finally, in Figure~\ref{fig:app:ot} we examine the coefficients of the Taylor series for the model dispersion relation developed in the main paper in Section~5.  We consider the case $Re=100$, $N=5$, and $\Lambda=-1.1$.   Panel~(a) refers to $\ar=2.1978$, and the singularities on the imaginary axis lie far from the point of interest $(\ar^*,0)$ (here $\ar^*$ refers to the real coordinate of the saddle point).  The first eight coefficients in the power series exhibit exponential decay.  On the other hand, in panel~(b) (for which $\ar=0.7739$), the coefficients decay algebraically.  This case corresponds to a $\ar\approx \ar^*$ (i.e. close to the saddle),  for which the radius of convergence is $R=0.80$; in the main part of the paper we extended finite truncations of the series beyond this point, i.e. $|\ai|>R$.  That the first eight coefficients of the power series decay algebraically  suggests that the  continuation described herein is appropriate.
Thus, it is possible (albeit with great caution) to use a finite truncation of a divergent power series to approximate the generating function of the power series.

\subsection{The series for the imaginary part of a function over-converges but the series for the real part does not}
\label{sec:roc:path}

The `pathological' example referred in the main part of Section~\ref{sec:roc} is the following:
\[
f(z)=\frac{1}{1-\imag z}=u(x,y)+\imag v(x,y),
\]
with a simple pole at $z=-\imag$.  The series representation
\[
f(z)=\frac{1}{1-\imag z},\qquad f_T(z)=\sum_{n=0}^\infty (\imag z)^n,\qquad f(z)=f_T(z)
\]
therefore converges on the interior of a disc of radius $R=1$ centred at $z=0$.  However, taking $z=\imag y$, we get
\[
v(0,y)=\Im\left(\frac{1}{1+y}\right)=0,\qquad
\Im\left(f_T(z)\right)=\Im\left[ \sum_{n=0}^\infty \left(-y\right)^n\right]=0.
\]
  Thus, the series for $v(0,y)$ is a trivial series whose coefficients are all zero, and which converges beyond the radius of convergence for all $y\neq -1$.  
The series $\Im(f_T(z))$ therefore agrees trivially with the function $v(0,y)$ for all $y\neq -1$.
However, at the same time,
\[
u(0,y)=\frac{1}{1+y},\qquad \Re(f_T(z))=\sum_{n=0}^\infty (-y)^n,
\]
and $u(0,y)=\Re(f_T(z))$ only for $|y|<R$.  Thus, we have an example where the imaginary part of $f(z)$ has a Taylor series that converges beyond the radius of convergence, on the line $\Re(z)=0$ with the point $\Im(z)=-1$ removed, i.e. the set $\{\Re(z)=0,\Im(z)\neq -1\}$.  However, the real part of $f(z)$ has a Taylor series that converges only inside the radius of convergence, along the same line.  Clearly, this is a very contrived example and for that reason, it was discussed in rather a dismissive fashion in the paper.

\subsection*{Acknowledgements}

L. \'O~N. would like to thank his  R. Smith for devising the example in Section~\ref{sec:roc:path} of this Supplementary Report.

\bibliographystyle{plainnat}